\documentclass{article}            %
\usepackage{amsmath}                %
\usepackage{amssymb}                %
\usepackage[english]{babel}             %
\def\la{\langle}
\def\ra{\rangle}
\def\trace{{\mathrm{tr}}}

\begin{document}
\small
\normalsize
\newcommand{\be}{\begin{equation}}
\newcommand{\ee}{\end{equation}}
\newcounter{saveeqn}
\newcommand{\alpheqn}{\setcounter{saveeqn}{\value{equation}}%
\setcounter{equation}{0}%
\renewcommand{\theequation}{\mbox{\arabic{saveeqn}-\alph{equation}}}}
\newcommand{\reseteqn}{\setcounter{equation}{\value{saveeqn}}%
\renewcommand{\theequation}{\arabic{equation}}}

\protect\newtheorem{principle}{Principle} %[section]
\protect\newtheorem{theo}[principle]{Theorem}
\protect\newtheorem{prop}[principle]{Proposition}
\protect\newtheorem{lem}[principle]{Lemma}
\protect\newtheorem{co}[principle]{Corollary}
\protect\newtheorem{de}[principle]{Definition}
\newtheorem{ex}[principle]{Example}
\newtheorem{rema}[principle]{Remark}
\newtheorem{state}[principle]{Statement}{\bf}{\rm}
\newtheorem{acknowledgements}[principle]{Acknowledgements}{\bf}{\rm}
\renewcommand{\baselinestretch}{1}
\small
\normalsize \vspace*{2.6cm} \noindent {\Large \textsf{The uniqueness
theorem for entanglement measures} } \\ \\ \\ \\ {\large {Matthew
J.~Donald}}  \\
\noindent {\normalsize The Cavendish
Laboratory, Madingley Road, Cambridge, CB3 0HE, Britain.
\\} {\normalsize E-mail: matthew.donald@phy.cam.ac.uk} \vspace*{0.6cm}
\noindent \\ {\large {Micha\l \/ Horodecki}}  \\
\noindent {\normalsize Institute of Theoretical Physics and
Astrophysics, University of Gda\'nsk, 80-952 Gda\'nsk, Poland.
\\} {\normalsize E-mail: fizmh@univ.gda.pl} \vspace*{0.6cm}
\noindent \\ {\large {Oliver Rudolph}}  \\ \noindent {\normalsize
Quantum Optics \& Information Group, Dipartimento di Fisica
``A.~Volta," Universit\`a di Pavia, Via Bassi 6, I-27100
Pavia, Italy. \\} {\normalsize E-mail: rudolph@fisicavolta.unipv.it}
\\
\\ \\ \\
\normalsize \noindent \emph{Abstract} We explore and develop the mathematics
of the theory of entanglement measures.  After a careful review and analysis
of definitions, of preliminary results, and of connections between
conditions on entanglement measures, we prove a sharpened version of a
uniqueness theorem which gives necessary and sufficient conditions for an
entanglement measure to coincide with the reduced von Neumann entropy on
pure states.  We also prove several versions of a theorem on extreme
entanglement measures in the case of mixed states.  We analyse properties of
the asymptotic regularization of entanglement measures proving, for example,
convexity for the entanglement cost and for the regularized relative entropy
of entanglement.
\\ \\
%Running title: \texttt{The uniqueness theorem for entanglement measures}

\section{Introduction}

Quantifying entanglement
\cite{BennettVSW97,PlenioV98,Vidal00,Horodecki99}
is one of the central topics of quantum
information theory. Any function that
quantifies entanglement is called an
entanglement measure.
Entanglement is a complex property of a state and, for arbitrary
states, there is no unique definitive measure. In general, there are
two ``regimes'' under which entanglement can be quantified: they may
be called the ``finite'' and the ``asymptotic'' regimes.  The first
deals with the entanglement of a single copy of a quantum state. In
the second, one is interested in how entanglement behaves when one
considers tensor products of a large number of identical copies of a
given state. It turns out that by studying the asymptotic regime it is
possible to obtain a clearer physical understanding of the nature of
entanglement.
This is seen, for example, in the so-called
``uniqueness theorem'' \cite{PopescuR97,Vidal00,Horodecki99, Nielsen-cont}
which states that, under appropriate conditions, all entanglement
measures coincide on pure bipartite
states and are equal to the von Neumann entropy of the corresponding
reduced density operator.
However, this theorem was never rigorously
proved under unified assumptions and definitions. Rather, there are
various versions of the argument scattered through the literature.

In Ref.~\cite{Horodecki99}, the uniqueness theorem
was put into a more general perspective.
Namely there are two basic measures of entanglement \cite{BennettVSW97} --
entanglement
of distillation ($E_D$) and entanglement cost ($E_C$)  --
 having the following
dual meanings:
\begin{itemize}
\item $E_D(\varrho)$ is the maximal number of singlets that can be produced
from the state $\varrho$ by means of local quantum
operations and classical
communication (\texttt{LQCC} operations).
\item $E_C(\varrho)$ is the minimal number of singlets needed to produce
the state $\varrho$ by \texttt{LQCC} operations.
\end{itemize}
(more precisely (cf.~Definitions \ref{distDef} and \ref{costDef}):
$E_D(\varrho)$ [$E_C(\varrho)$] is the maximal [minimal] number of singlets
{\it per copy} distillable from the state $\varrho$ [needed to form
$\varrho$] by
\texttt{LQCC} operations in the asymptotic regime of $n \rightarrow \infty$
copies). It is important here that the conversion is not required to be
perfect: the transformed state
needs to converge to the required state only in the limit of large $n$.
Now, in  Ref.~\cite{Horodecki99} it was shown that
the two basic measures of entanglement are,
respectively, a lower and an upper bound for any entanglement measure
satisfying appropriate
postulates in the asymptotic regime \cite{extreme-finite}. This suggests the
following clear picture: entanglement cost and
entanglement of distillation are extreme measures, and provided
they coincide on pure states, all other entanglement measures
coincide with them on pure states as well.
However as mentioned above, the
fact that $E_D$ and $E_C$ coincide on pure
states was not proven rigorously.  Moreover,
it turned out that the postulates
are too strong. They include convexity,  and some
additivity and continuity requirements.  It is not known whether any measure
exists which satisfies all the requirements.  $E_D$ and $E_C$ satisfy
the additivity requirement, but it is not known whether or not they are
continuous in the sense of Ref.~\cite{Horodecki99}.
There are also indications
that the entanglement of distillation is not convex
\cite{Shor2000}.
On the other hand, two other important measures, the
{\it entanglement of
formation} (denoted by $E_F$) and the
{\it relative entropy of entanglement} (denoted by $E_R$) are
continuous \cite{Nielsen-cont,DonaldH99} and convex, but there are
problems with additivity. The relative entropy of entanglement is
certainly not additive \cite{VollbrechtW00}, and we
do not know about the entanglement of formation.

In this situation it is desirable to prove the uniqueness theorem
from first principles, and to study to what extent we can
relax the
assumptions and still get uniqueness of entanglement measures
on pure states.
In the present paper
we have solved the problem completely
by providing {\it necessary and
sufficient} conditions for a measure of entanglement to be equal to
the von Neumann
entropy of the reduced density operator for pure states.
We also show that if we relax the postulate of
asymptotic continuity, then any measure of entanglement (not unique any
longer) for
pure states
must lie between the two analogues of $E_D$ and $E_C$ corresponding
to {\it perfect} fidelity of
conversion. These are $\tilde E_C (\psi)=S_0(\varrho)$ and
$\tilde E_D (\psi)=S_\infty(\varrho)$, where $\varrho$ is the reduced
density matrix of $\vert \psi \rangle$,
and $S_0$, $S_\infty$ are R\`enyi entropies.   In \cite{Rudolph01,Rudolph00},
one of us has studied entanglement measures based on cross norms and proved
an alternative uniqueness theorem for entanglement measures stemming from the
Khinchin-Faddeev characterization of Shannon entropy.

The present paper also contains further developments on
the problem of extreme measures. We
provide two useful new versions of the theorem of Ref.~\cite{Horodecki99}.
In one of them,
we show that for any (suitably normalized) function  $E$ for which the
regularization
$E^\infty(\varrho)=\lim_{n\to\infty}E(\varrho^{\otimes n})/n$ exists
and which is (i) nonincreasing
under local quantum operations and  classical
communication (\texttt{LQCC} operations) and (ii) asymptotically
continuous, the regularization $E^\infty$  must lie
between $E_D$ and $E_C$.  The theorem and its proof can easily be
generalized by replacing the class of
\texttt{LQCC} operations by
other classes of operations, or by considering conversions
between any two states.
Moreover, it is valid for multipartite cases. Therefore the result
will be an important tool for analysing asymptotic conversion
rates between different states.
In particular, it follows from our result that to
establish irreversibility of
conversion between two states (see \cite{Linden}),
one needs to compare regularizations
of asymptotically continuous entanglement measures for these states.

In the other new version of the extreme measures theorem, we are able to
weaken the postulates of Ref.~\cite{Horodecki99}, so that they are at least
satisfied by $E_C$. On the other hand, we do not have a proof that $E_D$ is
asymptotically continuous for mixed states, although there is strong
evidence that this is the case. If it is, then we would finally have a form
of the theorem, in which both $E_D$ and $E_C$ could be called extreme
measures, not only in the sense provided by the inequalities we prove, but
also in the sense that they belong to the set described by the postulates.

To obtain our results we perform a detailed study of possible
postulates for
entanglement measures in the finite and the asymptotic regime.
In particular, we examine which postulates
survive the operation of regularization. We show that if a function
is convex and subadditive (i.e., $f(\varrho\otimes \sigma) \leq
f(\varrho)+f(\sigma)$,)
then its regularization is convex too. Hence, both the
regularization of the relative entropy of
entanglement \cite{PlenioV98} as well as of the entanglement of
formation \cite{BennettVSW97} are convex.

It should be emphasized that our results are stated and proved
in language accessible for mathematicians or mathematical physicists
who have not previously been involved in quantum information theory. This is
in contrast to many papers in this field, where many implicit assumptions
are obstacles for understanding the meaning of the theorems and their
proofs  by non-specialists.  For this reason, we devote Sections
\ref{sec-preliminaries} and \ref{sec-classes} to careful statements of
some essential definitions and results. In Section \ref{sec-Nielsen} we
present a self-contained and straightforward proof of the difficult
implication in Nielsen's theorem. This is a theorem which we shall use
several times. Properties of entanglement measures and relations between
them are analysed in Section
\ref{sec-measures}.  The most prominent entanglement measures --
entanglement of  distillation, entanglement cost, entanglement of formation
and relative entropy of entanglement -- are
defined and studied in Section \ref{sec-examples}.
In Section \ref{sec-extreme} we present our versions of the theorem
on extreme measures. Finally,
Section \ref{sec-uniqueness} contains our version
of the uniqueness theorem for entanglement measures, stating
necessary and sufficient conditions for a functional to coincide
with the reduced von Neumann entropy on pure states.

%%%%%%%%%%%%%%%%%%%%%%%%%%%%%%%%%%%%%%%%%%%%%%%%%%%%%%%%%%%%%%%%%%%%%%

\section{Preliminaries}
%%%%%%%%%%%%%%%%%%%%%%%%%%%%%%%%%%%%%%%%%%%%%%%%%%%%%%%%%%%%%%%%%%%%%%

\label{sec-preliminaries}

Throughout this paper, all spaces considered are assumed to be
finite dimensional. The set of trace class operators on a
Hilbert space ${\mathcal{H}}$ is denoted by
${\mathcal{T}}({\mathcal{H}})$ and the set of bounded operators on
${\mathcal{H}}$ by ${\mathcal{B}}({\mathcal{H}})$. A density
operator (or \emph{state})
is a positive trace class operator with trace one.
The set of states on ${\mathcal{H}}$ is denoted by
${\Sigma}({\mathcal{H}})$ and the set of pure states by
${\Sigma_p}({\mathcal{H}})$. The trace class norm on
${\mathcal{T}}({\mathcal{H}})$ is denoted by $\Vert \cdot
\Vert_1$. For a wavefunction $\vert \psi \rangle \in {\mathcal{H}}$
the
corresponding state will be denoted by $P_\psi \equiv \vert \psi
\rangle \langle \psi \vert$. The \emph{support} of a trace class
operator is the subspace spanned by its eigenvectors with non-zero
eigenvalues.

In the present paper we restrict ourselves mainly to the situation of
a composite quantum system consisting of two subsystems with Hilbert
space
${\mathcal{H}}^A \otimes {\mathcal{H}}^B$ where ${\mathcal{H}}^A$ and
${\mathcal{H}}^B$ denote the Hilbert spaces of the subsystems.  Often
these systems are to be thought of as being spatially separate and
accessible to two independent observers, Alice and Bob.

\begin{de}
Let ${\mathcal{H}}^A$ and ${\mathcal{H}}^B$ be Hilbert spaces. A
density operator $\varrho$ on the tensor product
${\mathcal{H}}^A
\otimes {\mathcal{H}}^B$ is called \emph{separable} or
\emph{disentangled} if there exist a
sequence $(r_{i})$ of positive real numbers, a sequence
$(\rho^{A}_i)$ of density operators on
${\mathcal{H}}^A$ and a sequence $(\rho^{B}_i)$ of
density operators on ${\mathcal{H}}^B$ such that
\begin{equation} \label{e1}
\varrho = \sum_{i} r_{i} \rho^{A}_i \otimes \rho^{B}_i,
\end{equation}
where the sum converges in trace class norm.
\end{de}

The Schmidt decomposition \cite{Schmidt}
is of central importance in the
characterization and quantification of entanglement associated
with pure states.
\begin{lem} \label{Sch}
Let ${\mathcal{H}}^A$ and ${\mathcal{H}}^B$ be Hilbert spaces and let
$\vert \psi \rangle
\in {\mathcal{H}}^A \otimes {\mathcal{H}}^B$.
Then there exist a sequence of non-negative real numbers $(p_i)_i$ summing
to one and orthonormal bases $(\vert a_i \rangle)_i$ and $(\vert b_i
\rangle)_i$ of
${\mathcal{H}}^A$ and ${\mathcal{H}}^B$ respectively such that
\[ \vert \psi \rangle = \sum_i \sqrt{p_i} \vert a_i \otimes b_i
\rangle.
\]
{}\hfill $\blacksquare$
\end{lem}

By $S(\varrho)$ we will denote von Neumann entropy of the state
$\varrho$
given by
\begin{equation} \label{S}
S(\varrho):=-\trace\varrho\log_2\varrho.
\end{equation}

The \emph{von Neumann reduced entropy}
for a pure state $\sigma$ on a tensor product Hilbert space
${\mathcal{H}}^A \otimes {\mathcal{H}}^B$ is defined as
\begin{equation} \label{vN}
S_{\mathrm{vN}}(\sigma) := -
\trace_A((\trace_B\,\sigma) \log_2
(\trace_B\,\sigma)),
\end{equation} where
$\trace_A$ and $\trace_B$
denote the partial traces over
${\mathcal{H}}^A$ and ${\mathcal{H}}^B$ respectively. For $\sigma = P_\psi
= \vert \psi \rangle \langle \psi \vert$,
it is a straightforward consequence of Lemma \ref{Sch} that
\[ - \trace_A((\trace_B
P_\psi) \log_2 (\trace_B P_\psi)) = -
\trace_B((\trace_A P_\psi)
\log_2 (\trace_A P_\psi)) = - \sum_i p_i \log_2 p_i \]
where $( p_i )_i$ denotes the sequence of Schmidt coefficients of
$\vert \psi \rangle$.
However, for a general mixed state $\sigma$,
$\trace_A((\trace_B
\sigma)
\log_2 (\trace_B \sigma))$ may not equal
$\trace_B((\trace_A
\sigma) \log_2 (\trace_A \sigma))$.

%%%%%%%%%%%%%%%%%%%%%%%%%%%%%%%%%%%%%%%%%%%%%%%%%%%%%%%%%%%%%%%%%%

\section{Classes of quantum operations}
%%%%%%%%%%%%%%%%%%%%%%%%%%%%%%%%%%%%%%%%%%%%%%%%%%%%%%%%%%%%%%%%%%

\label{sec-classes}
In quantum information theory it is important to distinguish
between the class of quantum operations on a composite quantum
system which can be realized by separate \emph{local} actions on the
subsystems (i.e.~separate actions by ``Alice'' and by ``Bob'') and
those which cannot. The class of local quantum operations assisted by
classical communication (\texttt{LQCC}) is of central importance in
quantum cryptography and the emerging theory of quantum entanglement.

An \emph{operation} is a positive linear map $\Lambda
:{\mathcal{T}}({\mathcal{H}}_1) \to {\mathcal{T}}({\mathcal{H}}_2)$
such that $\trace(\Lambda(\sigma)) \leq 1$ for all $\sigma \in
\Sigma({\mathcal{H}}_1)$. \emph{Quantum operations} are operations which are
\emph{completely positive} \cite{Kraus71,Davies76}.  We shall
be interested in the trace preserving quantum operations. By the Choi-Kraus
representation
\cite{Kraus83,Kraus71,Davies76,Choi75}, these are precisely the
linear maps
$\Lambda :{\mathcal{T}}({\mathcal{H}}_1) \to {\mathcal{T}}({\mathcal{H}}_2)$
which can be written in the form
$\Lambda(B) = \sum_{i=1}^{n_1 n_2} W_i B W^\dagger_i$ for $B \in
{\mathcal{T}}({\mathcal{H}}_1)$
with operators
$W_i:{\mathcal{H}}_1 \to {\mathcal{H}}_2$ satisfying
$\sum_{i=1}^{n_1 n_2} W_i^\dagger W_i= \mathtt{1}_1$,
where $n_1 \equiv \dim {\mathcal{H}}_1$, $n_2 \equiv
\dim {\mathcal{H}}_2$, and $\mathtt{1}_1$ is the identity operator on
${\mathcal{H}}_1$. These can also be characterized as precisely the linear
maps which can be composed out of the following elementary operations
\begin{itemize}
\item[(O1)] Adding an uncorrelated ancilla: \\
$\Lambda_1 : {\mathcal{T}}({\mathcal{H}}_1) \to
{\mathcal{T}}({\mathcal{H}}_1
\otimes {\mathcal{K}}_1),
\Lambda_1(\rho) := \rho \otimes \sigma$, where ${\mathcal{H}}_1$ and
${\mathcal{K}}_1$ denote the Hilbert spaces of the original quantum
system and of the ancilla respectively and where $\sigma
\in \Sigma({\mathcal{K}}_1)$;
\item[(O2)] Tracing out part of the system: \\
$\Lambda_2 : {\mathcal{T}}({\mathcal{H}}_2 \otimes {\mathcal{K}}_2)
\to {\mathcal{T}}({\mathcal{H}}_2),
\Lambda_2(\rho) := \trace_{{\mathcal{K}}_2}(\rho)$ where
${\mathcal{H}}_2 \otimes {\mathcal{K}}_2$ and
${\mathcal{K}}_2$ denote the Hilbert spaces of the full original
quantum
system and of the dismissed part respectively and where
$\trace_{{\mathcal{K}}_2}$ denotes the partial trace over
${\mathcal{K}}_2$;
\item[(O3)] Unitary transformations: \\
$\Lambda_3 : {\mathcal{T}}({\mathcal{H}}_3) \to
{\mathcal{T}}({\mathcal{H}}_3), \Lambda_3(\rho) = U \rho U^\dagger$
where $U$ is a unitary operator on ${\mathcal{H}}_3$.
\end{itemize}

A discussion of this material with complete proofs from first
principles may be found in the initial archived draft of this
paper \cite{dhrv1}.

Defining a local operation as quantum operation on a individual subsystem, we
now turn to the definition of local operations assisted by classical
communication. As always in this paper we consider a quantum system
consisting of two (possibly separate) subsystems A and
B
with (initial) Hilbert spaces ${\mathcal{H}}^A$ and
${\mathcal{H}}^B$ respectively.
There are three cases: the communication between
A and B can be
unidirectional (in either direction) or bidirectional.

Let us first define the class of local quantum operations
({\texttt{LO}}) assisted by unidirectional
classical communication
(operations in this class will be called
one-way {\texttt{LQCC}} operations)
with direction from system A (Alice) to system
B (Bob).  In this case, the operations performed by
Bob depend on Alice's operations, but not conversely.

\begin{de}
A completely positive map $\Lambda: {\mathcal{T}}
({\mathcal{H}}^A_1\otimes{\mathcal{H}}^B_1) \to {\mathcal{T}}
({\mathcal{H}}^A_2\otimes{\mathcal{H}}^B_2)$
is called a {\textrm one-way {\texttt{LQCC}} operation} from A to B
if it can be written in the form
\begin{equation}
\Lambda(\sigma)=\sum_{i,j=1}^{K,L} ({\mathtt{1}}^A_2 \otimes
W^B_{ji})
(V^A_{i} \otimes {\mathtt{1}}^B_1) \sigma
(V^A_{i}{}^\dagger \otimes {\mathtt{1}}^B_1)({\mathtt{1}}^A_2
\otimes
W^B_{ji}{}^\dagger)
\end{equation} for all $\sigma \in {\mathcal{T}}({\mathcal{H}}^A_1
\otimes
{\mathcal{H}}^B_1)$ and some sequences of operators
$( V^A_{i} : {\mathcal{H}}^A_1 \to {\mathcal{H}}^A_2 )_{i}$ and
$(W^B_{ji} : {\mathcal{H}}^B_1 \to {\mathcal{H}}^B_2 )_{ji}$
with $\sum_{i=1}^K
V^A_{i}{}^\dagger V^A_{i}= {\mathtt{1}}^A_1$ and $\sum_{j=1}^L
W^B_{ji}{}^\dagger W^B_{ji}= {\mathtt{1}}^B_1$ for each $i$,
where ${\mathtt{1}}^A_1$, ${\mathtt{1}}^B_1$ and
${\mathtt{1}}^A_2$ are the unit operators acting on the
Hilbert spaces ${\mathcal{H}}^A_1$, ${\mathcal{H}}^B_1$ and
${\mathcal{H}}^A_2$,
respectively.
\end{de}

Of course, by the Choi-Kraus representation any operation $\Lambda$ of the
form
\begin{equation} \label{LO} \Lambda
= \Lambda^A \otimes I^B_1,
\end{equation} where $\Lambda^A : {\mathcal{T}}
({\mathcal{H}}^A_1) \rightarrow {\mathcal{T}}
({\mathcal{H}}^A_2)$ is a completely positive
trace preserving map  and $I_1^B$ is the identity operator on
${\mathcal{T}} ({\mathcal{H}}^B_1)$, is a one-way {\texttt{LQCC}}
operation from A to B.

Let us now define local quantum
operations assisted by bidirectional classical
communication ({\texttt{LQCC}} operations).

\begin{de}
A completely positive map $\Lambda: {\mathcal{T}}
({\mathcal{H}}^A\otimes{\mathcal{H}}^B)\rightarrow {\mathcal{T}}
({\mathcal{K}}^A\otimes{\mathcal{K}}^B)$
is called an {\texttt{LQCC}} \textrm{operation}
if there exist $n>0$ and sequences of Hilbert spaces
$({\mathcal{H}}^A_k)_{k=1}^{n+1}$ and
$({\mathcal{H}}^B_k)_{k=1}^{n+1}$ with
${\mathcal{H}}^{A(B)}_1 = {\mathcal{H}}^{A(B)}$ and
${\mathcal{H}}^{A(B)}_{n+1} = {\mathcal{K}}^{A(B)}$,
such that
$\Lambda$ can be written in the following form
\begin{equation}
\Lambda(\sigma) = \sum_{i_1,\ldots,i_{2n}=1}^{K_1,\ldots,K_{2n}}
V^{AB}_{i_1,\ldots,i_{2n}} \sigma
V^{AB}_{i_1,\ldots,i_{2n}}{}^\dagger
\end{equation} for all $\sigma \in {\mathcal{T}}({\mathcal{H}}^A
\otimes {\mathcal{H}}^B)$
where $V^{AB}_{i_1,\ldots,i_{2n}} :
{\mathcal{H}}^A\otimes{\mathcal{H}}^B
\to {\mathcal{K}}^A\otimes{\mathcal{K}}^B$ is given by
\[
V^{AB}_{i_1,\ldots,i_{2n}} :=
({\mathtt{1}}_{n+1}^A
\otimes W_{2n}^{i_{2n}, \ldots, i_1})(V_{2n - 1}^{i_{2n - 1},\ldots, i_1}
\otimes {\mathtt{1}}_n^B) ({\mathtt{1}}_{n}^A
\otimes W_{2n - 2}^{i_{2n - 2}, \ldots, i_1})
\cdots ( {\mathtt{1}}_2^A \otimes W_{2}^{i_{2},i_{1}})
(V^{i_{1}}_{1} \otimes {\mathtt{1}}_{1}^B)
\]
with families of operators
\addtocounter{equation}{1}
\alpheqn
\begin{eqnarray}
&& \left( V_{2k - 1}^{i_{2k-1}, \ldots, i_1} : {\mathcal{H}}^A_{k}
\to {\mathcal{H}}^A_{k+1} \right)_{k=1}^{n}, \\
&& \left( W_{2k}^{i_{2k}, \ldots, i_1} : {\mathcal{H}}^B_{k}
\to {\mathcal{H}}^B_{k+1} \right)_{k=1}^n \end{eqnarray}
\reseteqn
\addtocounter{equation}{1}
\alpheqn
such that for
$k = 0, \dots, n-1$ and each sequence of indices $(i_{2k}, \ldots, i_1)$
\begin{equation} \sum_{i_{2k+1}=1}^{K_{2k+1}} (V_{2k +
1}^{i_{2k+1},\ldots, i_1})^\dagger V_{2k + 1}^{i_{2k+1},\ldots, i_1} =
{\mathtt{1}}^A_{k+1}
\end{equation}
and for $k = 1, \dots, n$ and each sequence of indices $(i_{2k-1}, \ldots,
i_1)$
\begin{equation} \sum_{i_{2k}=1}^{K_{2k}} (W_{2k}^{i_{2k},\ldots,
i_1})^\dagger W_{2k}^{i_{2k},\ldots, i_1} = {\mathtt{1}}^B_{k}
\end{equation}
\reseteqn
where for all $k>0$, \,
${\mathtt{1}}^A_k$ and ${\mathtt{1}}^B_k$ denote the unit
operator on ${\mathcal{H}}^A_k$ and ${\mathcal{H}}^B_k$
respectively.
\end{de}

Obviously the class of one-way \texttt{LQCC} operations is
a subclass of the class of \texttt{LQCC} operations.
There is another important class:
separable operations. A separable operation is an operation
of the form:
\begin{equation}
\Lambda : {\mathcal{T}}({\mathcal{H}}^A \otimes {\mathcal{H}}^B)
\to {\mathcal{T}}({\mathcal{K}}^A \otimes {\mathcal{K}}^B), \quad
\Lambda(\sigma) \equiv
\sum_{i=1}^k (V_i\otimes W_i)\sigma (V_i\otimes W_i)^\dagger
\end{equation}
with $\sum_{i=1} (V_i\otimes W_i)^\dagger V_i\otimes W_i=
{\mathtt{1}}^{AB}$ where ${\mathtt{1}}^{AB}$ denotes the unit
operator
acting on ${\mathcal{H}}^A \otimes {\mathcal{H}}^B$.
The class of separable operations
is strictly larger than the {\texttt{LQCC}} class
\cite{nonlocality}.

One can also consider a small class obtained by taking the convex hull
${\mathtt{C}}$ of the set of all maps of the form
$\Lambda^A\otimes \Lambda^B$. Such operations require in general
one-way classical communication, but they do not cover the
whole class of
one-way {\texttt{LQCC}} operations.

All the classes above are closed under tensor multiplication, convex
combinations, and composition. The results of our paper apply
in principle
to all the classes apart from the last (i.e., apart from the class
of
all operations in the convex hull $\mathtt{C}$ of the set of
all maps of the form
$\Lambda^A\otimes \Lambda^B$). For definiteness, in the sequel we
will use {\texttt{LQCC}} operations.

Finally, we conclude this section with a useful technical lemma.

\begin{lem} \label{cont} Let
$\Lambda : {\mathcal{T}}({\mathcal{H}}_1) \to
{\mathcal{T}}({\mathcal{H}}_2)$ be a positive trace-preserving map
and
suppose that $B \in {\mathcal{T}}({\mathcal{H}}_1)$ with $B = B^*$.
Then
$\Vert \Lambda(B) \Vert_1 \leq \Vert B \Vert_1$.
\end{lem}

\noindent \textbf{Proof}:  Suppose that $B$ has eigenvalue expansion
$B = \sum_{i=1}^{n_1} \beta_i |\psi_i\rangle \langle \psi_i|$.
Then
\[ \Vert \Lambda(B) \Vert_1 \leq \sum_{i=1}^{n_1} |\beta_i|
\left\Vert \Lambda(|\psi_i\rangle \langle \psi_i|) \right \Vert_1
= \Vert B \Vert_1 \] as $\Vert B \Vert_1 = \sum_{i=1}^{n_1}
|\beta_i|$
and
$\Lambda(|\psi_i\rangle \langle \psi_i|)$ is a positive trace class
operator with unit trace. \hfill $\blacksquare$

%%%%%%%%%%%%%%%%%%%%%%%%%%%%%%%%%%%%%%%%%%%%%%%%%%%%%%%%%%%%%%%%%%%

\section{Nielsen's theorem}

%%%%%%%%%%%%%%%%%%%%%%%%%%%%%%%%%%%%%%%%%%%%%%%%%%%%%%%%%%%%%%%%%%%

\label{sec-Nielsen}
A beautiful and powerful result of entanglement theory is
Nielsen's theorem \cite{Nielsen99}.   In one direction, the
proof is straightforward, and we refer to \cite{Nielsen99}.  The
other direction is more difficult.  We present here an entirely
self-contained, simple, and direct proof.  Alternative proofs
have previously been given by Hardy \cite{Har99}
and by Jensen and Schack \cite{JenSch00}.

Before we state the theorem we need the
following definition.
\begin{de} Let
$(p_i)_{i=1}^{m_1}$ and
$(q_i)_{i=1}^{m_2}$ be two probability distributions
with probabilities arranged in decreasing order,
i.e.,
$p_1\geq p_2\geq \ldots \geq p_{m_1}$ and similarly for $(q_i)_i$.
Then we will say that $(q_i)_i$ \emph{majorizes} $(p_i)_i$ (in
symbols
$(q_i)_i \succ (p_i)_i$) if for all $k\leq\min\{m_1,m_2\}$
we have
\begin{equation}
\sum_{i=1}^k q_i\geq \sum_{i=1}^k p_i.
\end{equation} \end{de}

\begin{theo}[Nielsen] Let ${\mathcal{H}}^A$ and ${\mathcal{H}}^B$ be
Hilbert spaces
and let $( \vert \chi_m \rangle )_{m=1}^M$
and $( \vert \kappa_m \rangle )_{m=1}^M$ be
orthonormal bases for ${\mathcal{H}}^A$ and
${\mathcal{H}}^B$ respectively.
Let $\vert \Psi \rangle = \sum_{m=1}^M \sqrt{p_m} |\chi_m \kappa_m
\rangle $ and
$\vert \Phi \rangle =
\sum_{m=1}^M \sqrt{q_m} |\chi_m \kappa_m\rangle $ be
Schmidt decompositions of normalized vectors $\vert \Psi \rangle$
and $\vert \Phi \rangle$ in ${\mathcal{H}}^A \otimes
{\mathcal{H}}^B$
with $p_1 \geq p_2 \geq \dots \geq p_M$ and $q_1 \geq q_2 \geq
\dots \geq q_M$.
Then $\vert \Psi \rangle\langle \Psi \vert$ can be converted into
$\vert
\Phi \rangle\langle \Phi \vert$ by {\texttt{LQCC}}
operations if and only if $(q_i)$ majorises $(p_i)$.
\label{Nielsen}
\end{theo}

\noindent \textbf{Proof}: (One direction only.)  Suppose that $(q_i)$
majorises
$(p_i)$. Set $\rho \equiv
|\Psi\rangle \langle \Psi|$ and $\sigma \equiv |\Phi\rangle
\langle \Phi|$.
We shall prove that
there is a sequence $(\Lambda_n )_{n=1}^{N}$ with $N < M$ of
completely
positive maps on ${\mathcal{T}}({\mathcal{H}}^A
\otimes {\mathcal{H}}^B)$ of the form
\begin{equation} \label{Nfrm} \Lambda_n(\omega)  = (C_{n} \otimes
U_{n}) \omega (C_{n} \otimes U_{n})^\dagger + (D_{n} \otimes V_{n})
\omega (D_{n} \otimes V_{n})^\dagger \end{equation}
where $U_{n}, V_{n} \in {\mathcal{B}}({\mathcal{H}}^B)$ are unitary
and
$C_{n}, D_{n} \in {\mathcal{B}}({\mathcal{H}}^A)$
satisfy $C_{n}^\dagger C_{n} +
D_{n}^\dagger D_{n} = {\mathtt{1}}^A$ such that
$\Lambda_1 \circ \Lambda_2 \circ \cdots \circ \Lambda_N(\rho) =
\sigma$.
Note that all the $\Lambda_n$ are one-way
{\texttt{LQCC}} operations from A to B and
hence their composition also is.  As the Schmidt decomposition is
symmetrical between A and B, we could also use one-way
{\texttt{LQCC}}
operations from B to A. Set $\delta_k \equiv \sum_{m = 1}^k q_m -
\sum_{m
= 1}^k p_m$ for $k = 1, 2, \cdots, M$.
Then $\delta_M = 0$.
Let $N = N(\vert \Psi \rangle, \vert \Phi \rangle)$
be the number of non-zero $\delta_k$.
We shall prove the result by induction on $N$.
$\vert \Psi \rangle = \vert \Phi \rangle$
if and only if $\delta_1 = \delta_2 = \cdots = \delta_{M-1} =
0$. In this case $N(\vert \Psi \rangle, \vert \Phi \rangle) = 0$,
$\rho = \sigma$, and the result is certainly true.

Suppose that the result holds for all pairs $(\vert \Psi \rangle,
\vert \Phi \rangle)$ satisfying the
conditions of the proposition with $N( \vert \Psi \rangle, \vert
\Phi
\rangle) = 0, \cdots, L$ and that $(\vert \Psi \rangle, \vert
\Phi \rangle)$ is a pair with $N(\vert \Psi \rangle, \vert \Phi
\rangle) = L + 1$.
Then there exists $J \geq 1$ such that $\delta_1 = \delta_2 = \cdots
=
\delta_{J-1} = 0$ and $\delta_J > 0$.
Setting $\delta_0 := 0$, we have
$q_j - p_j = \delta_{j-1} + q_j - p_j = \delta_j$ for
$j = 1, \cdots, J$.
This implies that $p_j = q_j$ for $j = 1, \cdots, J-1$ and that $q_J
>
p_J$.
Suppose that $\delta_k > 0$ for $k = J, J+1, \cdots, K-1$ and that
$\delta_{K} = 0$.
$p_K - q_K = p_K - q_K + \delta_{K} = \delta_{K-1}$
and $p_K > q_K$.
Moreover, if $K < M$ then $q_{K+1} - p_{K+1} = \delta_K + q_{K+1} - p_{K+1}
=
\delta_{K+1}
\geq 0$.
Summarizing, we have
$$p_{J-1} = q_{J-1} \geq q_J > p_J \geq p_K > q_K \geq q_{K+1} \geq
p_{K+1}.$$

Define $(r_m)_{m=1}^M$ by $r_m := p_m$ for $m \neq J, K$ and
by $r_J := p_J + \delta$, $r_K := p_K - \delta$ where
$\delta := \min\{\delta_k : k = J, \cdots, K - 1\}$.
By construction $\delta > 0$.
Now $\delta \leq \delta_J$ implies $q_J \geq r_J \geq p_J$
and $\delta \leq
\delta_{K-1}$ implies $p_K \geq r_K \geq
q_K$.  This in turn implies that
$r_1 \geq r_2 \geq \cdots \geq r_M$.
Thus for $k = 1, \cdots, J-1$ and for $k = K, \cdots, M$,
$$\textstyle\sum_{m = 1}^k r_m = \sum_{m = 1}^k p_m \leq
\sum_{m = 1}^k q_m.$$
For $k = J, \dots, K-1$,
$\sum_{m = 1}^k r_m = \sum_{m = 1}^k p_m + \delta$ and so,
as $0 < \delta \leq \delta_k$,
$$\textstyle \sum_{m = 1}^k p_m < \sum_{m = 1}^k r_m \leq
\sum_{m = 1}^k q_m.$$
Define $\vert \Xi \rangle :=
\sum_{m=1}^M \sqrt{r_m} |\chi_m \kappa_m \rangle$.
Then $N(\vert \Xi \rangle, \vert \Phi \rangle)
\leq L$ so that, by the inductive hypothesis,
there is a sequence $(\Lambda_n)_{n=1}^{N}$ of maps of the
required form with $N = N(\vert \Xi \rangle, \vert \Phi \rangle)$
such that $$\Lambda_1 \circ \Lambda_2 \circ \cdots
\circ \Lambda_N (| \Xi \rangle \langle \Xi |) =
\sigma.$$
Thus to complete the proof, we need only find a completely positive
map
$\Lambda$ of the required form such that
\begin{equation} \label{eqNiel}\Lambda(|\Psi\rangle \langle \Psi|) =
|\Xi\rangle \langle \Xi|.\end{equation}

To this end set $P := \sum_{m \neq J, K} |\chi_m \rangle \langle
\chi_m |$.  Set
\begin{equation}
C := \sqrt{ \frac{r_J p_J - r_K p_K}{r^2_J - r^2_K}} \left( P
+  \sqrt{\frac{r_J}{p_J}} \left| \chi_J \right\rangle \left\langle
\chi_J \right| +
\sqrt{\frac{r_K}{p_K}} \left| \chi_K \right\rangle \left\langle
\chi_K \right| \right)
\nonumber \end{equation} and $U := {\mathtt{1}}^B$.
Set \begin{eqnarray*}
D & := & \sqrt{\frac{r_J p_K - r_K p_J}{r^2_J - r^2_K}} \left( P
+ \sqrt{\frac{r_K}{p_J}} \left| \chi_K \right\rangle \left\langle
\chi_J \right| + \sqrt{\frac{r_J}{p_K}} \left|
\chi_J \right\rangle \left\langle \chi_K \right| \right) \\
\mbox{and  } V & := & \left| \kappa_K \right\rangle \left\langle
\kappa_J \right| +
\left| \kappa_K \right\rangle \left\langle \kappa_J \right| +
\sum_{m \neq J,K} \left| \kappa_m \right\rangle \left\langle
\kappa_m \right|.
\end{eqnarray*}
Note that $p_J \geq p_K > q_K \geq 0$, that $r_J > r_K$,
that $r_J p_J > r_K p_K$, and that
$r_J p_K - r_K p_J = (p_J + \delta) p_K - (p_K - \delta) p_J =
\delta (p_K + p_J) > 0$.
Note also that $r^2_J - r^2_K = (r_J - r_K)(r_J + r_K) =
(r_J - r_K)(p_J + p_K)$ so that
\[ \frac{r_J p_J - r_K p_K}{r^2_J - r^2_K} +
\frac{r_J p_K - r_K p_J}{r^2_J - r^2_K} = 1. \]

With these definitions and notes, the completion of the proof is
straightforward.
\hfill $\blacksquare$

%%%%%%%%%%%%%%%%%%%%%%%%%%%%%%%%%%%%%%%%%%%%%%%%%%%%%%%%%%%%%%%%%%%%

\section{Entanglement measures}

%%%%%%%%%%%%%%%%%%%%%%%%%%%%%%%%%%%%%%%%%%%%%%%%%%%%%%%%%%%%%%%%%%%%

\begin{de}
Consider a bipartite composite quantum system with Hilbert space of
the form ${\mathcal{H}}^A \otimes {\mathcal{H}}^B$ where
${\mathcal{H}}^A \equiv {\mathcal{H}}^B \equiv {\mathbb{C}}^d$.
Assume
that isomorphisms between  ${\mathbb{C}}^d$,
${\mathcal{H}}^A$, and ${\mathcal{H}}^B$ are chosen.
For a chosen orthonormal basis $(
\vert \psi_i \rangle )_{i=1}^d$ of ${\mathbb{C}}^d$,
we let
\[ \vert \Psi_+({\mathbb{C}}^d) \rangle \equiv
\sum_{i=1}^d {1
\over \sqrt{d}} |\psi_i \otimes \psi_i \rangle \in
{\mathcal{H}}^A \otimes {\mathcal{H}}^B. \]
$\vert \Psi_+({\mathbb{C}}^d) \rangle$ is a \emph{maximally
entangled} wavefunction. All other maximally entangled wavefunctions
in
${\mathcal{H}}^A \otimes {\mathcal{H}}^B$ can be obtained by
applying
a unitary operator of the form $\mathtt{1}^A
\otimes U^B$ to $\vert \Psi_+({\mathbb{C}}^d) \rangle$ where $U^B$
is
a unitary operator on ${\mathcal{H}^B}$.
The pure state  corresponding to $\vert \Psi_+({\mathbb{C}}^d)
\rangle$
will be denoted  by
$P_+({\mathbb{C}}^d) \equiv \vert \Psi_+({\mathbb{C}}^d) \rangle
\langle \Psi_+({\mathbb{C}}^d) \vert$.

In an arbitrary bipartite composite system,
we shall refer to any wavefunction with the same Schmidt
coefficients as $\vert \Psi_+({\mathbb{C}}^d)\rangle$ as a
\emph{representative} of
$\vert \Psi_+({\mathbb{C}}^d)\rangle$  and to the corresponding
state as a \emph{representative} of
$P_+({\mathbb{C}}^d)$.
\end{de}

\label{sec-measures}
\subsection{Conditions on mixed states}
The degree of entanglement of a density operator on the Hilbert
space
of a bipartite composite quantum system can be expressed by an
``entanglement measure.''  This a non-negative real-valued
functional
$E$
defined on $\Sigma({\mathcal{H}}^A \otimes {\mathcal{H}}^B)$ for all
finite-dimensional Hilbert
spaces ${\mathcal{H}}^A$ and ${\mathcal{H}}^B$.
Any of the following conditions might be imposed on $E$
\cite{BennettVSW97,VedralPRK97,PlenioV98,Horodecki99,Vidal00}
\begin{itemize}
\item[(E0)] If $\sigma$ is separable, then $E(\sigma)
= 0$.
\item[(E1)] (Normalization.) If $P_+^d$ is any representative of
$P_+({\mathbb{C}}^d)$, then
$E(P_+^d) =
\log_2 d$ for $d = 1, 2, \dots$.
\end{itemize}
A weaker condition is:
\begin{itemize}
\item[(E1$'$)] $E(P_+({\mathbb{C}}^2)) = 1$.
\item[(E2)] ({\texttt{LQCC}} Monotonicity.) Entanglement cannot
increase under
procedures consisting of
local operations on the two quantum systems and classical
communication. If $\Lambda$ is an \texttt{LQCC}
operation, then
\begin{equation}
E(\Lambda(\sigma)) \leq E(\sigma) \label{LQCC}
\end{equation}
for all $\sigma \in \Sigma({\mathcal{H}}^A
\otimes {\mathcal{H}}^B)$.
\end{itemize}
A condition which, as we shall confirm below (Lemma \ref{lemE2}), is
weaker than (E2), is
\begin{itemize}
\item[(E2$'$)]  $E(\Lambda(\sigma)) = E(\sigma)$ whenever  $\sigma
\in \Sigma({\mathcal{H}}^A \otimes {\mathcal{H}}^B)$ and $\Lambda$
is
a strictly local operation which is either unitary or which adds
extraneous dimensions.  On Alice's side, these local operations take
the form, either of
$\Lambda_1(\varrho) = (U^A \otimes I^B) \varrho (U^A \otimes
I^B)^\dagger$ where $U^A: {\mathcal{H}}^A \to {\mathcal{H}}^A$ is
unitary, or of $\Lambda_2(\varrho) = (W^A \otimes I^B) \varrho (W^A
\otimes I^B)^\dagger$ where ${\mathcal{H}}^A \subset
{\mathcal{K}}^A$
and $W^A: {\mathcal{H}}^A \to {\mathcal{K}}^A$ is the inclusion map.
There are equivalent local operations on Bob's side.
 \item[(E2$''$)]  $E(\Lambda(\sigma)) = E(\sigma)$ whenever
$\sigma
\in \Sigma({\mathcal{H}}^A \otimes {\mathcal{H}}^B)$ and $\Lambda$
is
a strictly local unitary operation.
\end{itemize}
Without further remark, we shall always assume that all our measures
satisfy (E2$''$).
\begin{itemize}
\item[(E3)] (Continuity.)
Let $({\mathcal{H}}^A_n )_{n \in {\mathbb{N}}}$ and
$({\mathcal{H}}^B_n )_{n \in {\mathbb{N}}}$ be sequences of
Hilbert spaces and let ${\mathcal{H}}_n \equiv {\mathcal{H}}^A_n
\otimes {\mathcal{H}}^B_n$ for all $n$.
For all sequences $(\varrho_n)_{n \in {\mathbb{N}}}$
and $(\sigma_n)_{n \in {\mathbb{N}}}$ of states
with $\varrho_n, \sigma_n \in {\Sigma}({\mathcal{H}}^A_n
\otimes {\mathcal{H}}^B_n)$, such that
$ \Vert \, \varrho_n - \sigma_n \Vert_1 \to 0$, we require that
\[
\frac{E(\varrho_n)-E(\sigma_n)}{1 + \log_2 \dim{\mathcal{H}}_n} \to
0.
\]
\end{itemize}
A weaker condition deals only with approximations to pure states:
\begin{itemize}
\item[(E3$'$)] Same as (E3) but with
$\varrho_n\in\Sigma_p({\mathcal{H}}^A_n
\otimes {\mathcal{H}}^B_n)$ for all $n$.
\end{itemize}
Sometimes we are interested in entanglement measures which satisfy an
additivity property:
\begin{itemize}
\item[(E4)]  (Additivity.) For all $n \geq 1$ and all
$\varrho \in
\Sigma({\mathcal{H}}^A \otimes {\mathcal{H}}^B)$
\[ {E(\varrho^{\otimes n})\over n} = E(\varrho). \]
\end{itemize}
Here
$\varrho^{\otimes n}$ denotes the $n$-fold tensor product of
$\varrho$ by itself which acts on the tensor product
$({\mathcal{H}}^A)^{\otimes n} \otimes ({\mathcal{H}}^B)^{\otimes
n}$.

An apparently weaker property, which as we shall see in Lemma
\ref{lemadd} is actually equivalent to (E4), is
\begin{itemize}
\item[(E4$'$)]  (Asymptotic Additivity.) Given $\epsilon > 0$ and
$\varrho \in \Sigma({\mathcal{H}}^A \otimes {\mathcal{H}}^B)$,
there exists an integer $N>0$ such that $n \geq N$ implies
\[ {E(\varrho^{\otimes n})\over n} - \epsilon
\leq E(\varrho) \leq {E(\varrho^{\otimes n})\over n} + \epsilon. \]
\item[(E5)]  (Subadditivity.) For all
$\varrho, \sigma \in \Sigma({\mathcal{H}}^A \otimes
{\mathcal{H}}^B)$,
\[ E(\varrho \otimes \sigma) \leq E(\varrho) + E(\sigma). \]
\end{itemize}
\begin{itemize}
\item[(E5$'$)]  For all
$\varrho \in \Sigma({\mathcal{H}}^A \otimes {\mathcal{H}}^B)$
and $m, n \geq 1$,
\[ E\left(\varrho^{\otimes (m + n)} \right)
\leq E\left(\varrho^{\otimes m} \right) + E\left(\varrho^{\otimes n}
\right). \]
\item[(E5$''$)]  (Existence of a regularization.) For all
$\varrho \in \Sigma({\mathcal{H}}^A \otimes {\mathcal{H}}^B)$,
the limit
\[
E^{\infty}(\varrho)\equiv \lim_{n\rightarrow\infty}
{E\left(\varrho^{\otimes n} \right)\over
n} \]
exists.
\end{itemize}
In Lemma \ref{lemE5} we shall prove the well-known result that
(E5$'$)
is a sufficient condition for (E5$''$). When (E5$''$) holds,
we shall refer to $E^{\infty}$ as the regularization of $E$.  We
shall
discuss some general properties of $E^\infty$ in Proposition
\ref{reg}.
\begin{itemize}
\item[(E6)] (Convexity.) Mixing of states does not increase
entanglement. \[ E(\lambda \varrho + (1 - \lambda) \sigma) \leq
\lambda E(\varrho) + (1 - \lambda) E(\sigma) \] for all $0 \leq
\lambda \leq 1$ and all $\varrho, \sigma \in \Sigma({\mathcal{H}}^A
\otimes {\mathcal{H}}^B)$.
\end{itemize}
(E6) might seem to be essential for a measure of
entanglement.  Nevertheless, there is some evidence that an
important
entanglement measure (the entanglement of distillation)
which describes asymptotic properties of multiple copies of
identical
states may not be convex \cite{Shor2000}.
A weaker condition is to require convexity
only on decompositions into pure states.  We shall prove below that
this property is satisfied by the entanglement of distillation.
\begin{itemize}
\item[(E6$'$)] For any state  $\varrho \in \Sigma({\mathcal{H}}^A
\otimes {\mathcal{H}}^B)$ and any decomposition
$\varrho=\sum_ip_i |\psi_i\rangle\langle\psi_i|$
with $\vert \psi_i \rangle \in {\mathcal{H}}^A \otimes
{\mathcal{H}}^B$, $p_i \geq 0$ for all $i$ and $\sum_ip_i=1$,
we require
\[
E(\varrho)\leq \sum_ip_i E(P_{\psi_i}).
\]
\end{itemize}

%%%%%%%%%%%%%%%%%%%%%%%%%%%%%%%%%%%%%%%%%%%%%%%%%%%%%%%%%%%%

\subsection{Conditions on pure states}

%%%%%%%%%%%%%%%%%%%%%%%%%%%%%%%%%%%%%%%%%%%%%%%%%%%%%%%%%%%%

The conditions imposed on an entanglement measure can be weakened
by requiring that they only apply for pure states.  Indeed,
it might not even be required that the measure is defined except on
pure states.  Recall that ${\Sigma_p}({\mathcal{H}}^A \otimes
{\mathcal{H}}^B)$ denotes the set of pure states on the composite
space.

\begin{itemize}
\item[(P0)] If $\sigma \in {\Sigma_p}({\mathcal{H}}^A \otimes
{\mathcal{H}}^B)$ is separable, then $E(\sigma)
= 0$.
\item[(P1)]  = (E1) (Normalization.) If $P_+^d$ is any
representative
of $P_+({\mathbb{C}}^d)$, then
$E(P_+^d) = \log_2 d$ for $d = 1, 2, \dots$.
\item[(P1$'$)] = (E1$'$) $E(P_+({\mathbb{C}}^2)) = 1$.
\item[(P2)] Let $\Lambda$ be an operation which can be
realized by means of local operations and classical
communications. If  $\sigma \in {\Sigma_p}({\mathcal{H}}^A \otimes
{\mathcal{H}}^B)$ is such that $\Lambda(\sigma)$ is also pure, then
\[
E(\Lambda(\sigma)) \leq E(\sigma). \]
\item[(P2$'$)] For $\sigma \in {\Sigma_p}({\mathcal{H}}^A \otimes
{\mathcal{H}}^B)$, $E(\sigma)$ depends only on the non-zero
coefficients of a Schmidt decomposition of $\sigma$.
\end{itemize}
By Nielsen's theorem and the proof of Lemma \ref{lemE2} below,
(P2) is equivalent to assuming (P2$'$) and that if the Schmidt
coefficients of $\varrho$ majorize those of $\sigma$ then $E(\varrho) \leq
E(\sigma)$.  Our proof of the theorem shows that, given (P2$'$), only
local operations and operations of the specific form of
Equation (\ref{Nfrm})
need be considered for (P2) (cf. \cite{Joathan}).

Below we will in particular be interested in
entanglement measures satisfying the following additional
conditions:
\begin{itemize}
\item[(P3)]
Let $({\mathcal{H}}^A_n )_{n \in {\mathbb{N}}}$ and
$({\mathcal{H}}^B_n )_{n \in {\mathbb{N}}}$ be sequences of
Hilbert spaces  and let ${\mathcal{H}}_n \equiv {\mathcal{H}}^A_n
\otimes {\mathcal{H}}^B_n$ for all $n$.
For all sequences $(\varrho_n)_{n \in {\mathbb{N}}}$
and $(\sigma_n)_{n \in {\mathbb{N}}}$ of states
with $\varrho_n, \sigma_n \in {\Sigma}_p({\mathcal{H}}^A_n
\otimes {\mathcal{H}}^B_n)$, such that
$ \Vert \, \varrho_n - \sigma_n \Vert_1 \to 0$, we require that
\[
\frac{E(\varrho_n)-E(\sigma_n)}{1 + \log_2 \dim{\mathcal{H}}_n} \to
0.
\]
\item[(P4)]  For all $n \geq 1$ and all
$\varrho \in
\Sigma_p({\mathcal{H}}^A \otimes {\mathcal{H}}^B)$,
\[ {E(\varrho^{\otimes n})\over n} = E(\varrho). \]
\end{itemize}
Of course, when $\varrho$ is pure, so is $\varrho^{\otimes n}$.
\begin{itemize}
\item[(P4$'$)]  Given $\epsilon > 0$ and
$\varrho \in \Sigma_p({\mathcal{H}}^A \otimes {\mathcal{H}}^B)$,
there exists an integer $N>0$ such that $n \geq N$ implies
\[ {E(\varrho^{\otimes n})\over n} - \epsilon
\leq E(\varrho) \leq {E(\varrho^{\otimes n})\over n} + \epsilon. \]
\item[(P5$''$)]  (Existence of a regularization on pure states.) For all
$\varrho \in \Sigma_p({\mathcal{H}}^A \otimes {\mathcal{H}}^B)$,
the limit
\[
E^{\infty}(\varrho)\equiv \lim_{n\rightarrow\infty}
{E\left(\varrho^{\otimes n} \right)\over
n} \]
exists.
\end{itemize}

%%%%%%%%%%%%%%%%%%%%%%%%%%%%%%%%%%%%%%%%%%%%%%%%%%%%%%%%%%%%

\subsection{Some connections between the conditions}

%%%%%%%%%%%%%%%%%%%%%%%%%%%%%%%%%%%%%%%%%%%%%%%%%%%%%%%%%%%%

\begin{lem}
\label{lemE2} (E2$'$) is implied by (E2).
\end{lem}
{\bf Proof:}  By Equation (\ref{LO}), the operations considered in
(E2$'$) are \texttt{LQCC}.  To see this for $\Lambda_2$, note that
$W^A{}^\dagger W^A = 1_{{\mathcal{H}}^A}$. Thus (E2) implies
$E(\Lambda_i(\sigma)) \leq E(\sigma)$ for $i = 1, 2$.  Unitary
maps are invertible and so $E(\Lambda_1(\sigma)) \geq E(\sigma)$.
On
the other hand, if ${\mathcal{H}}^A \subset
{\mathcal{K}}^A$ and $P^A$ is the projection onto ${\mathcal{H}}^A$,
then, for any $\tau^A \in \Sigma_p({\mathcal{H}}^A)$, the map
$\Lambda^A_3 : \Sigma({\mathcal{K}}^A) \to
\Sigma({\mathcal{H}}^A)$ defined by
$\Lambda_3^A(\varrho) := P^A \varrho P^A{}^\dagger +
{\mathrm{tr}}(\varrho(1 - P^A))
\tau^A $ is completely positive and trace preserving, so by
Equation (\ref{LO}), the map on $\Sigma({\mathcal{K}}^A \otimes
{\mathcal{H}}^A)$ defined by
$\Lambda_3 = \Lambda^A_3 \otimes I^B$ is \texttt{LQCC}.

$\Lambda_3(\Lambda_2(\sigma)) = \sigma$ and hence (E2) implies
$E(\sigma) \leq E(\Lambda_2(\sigma))$. \hfill $\blacksquare$

\begin{lem}
\label{lemadd} (E4$'$) is equivalent to (E4) and (P4$'$) is
equivalent
to (P4).
\end{lem}
{\bf Proof:} That (E4) implies (E4$'$) is immediate.  Suppose
(E4$'$)
and choose $m$, $\varrho$, and $\epsilon$.

By (E4$'$), there exists
$N$ such that $n \geq N$ implies $|E(\varrho) -
{E(\varrho^{\otimes n})/ n}| \leq \epsilon$ and
$|E(\varrho^{\otimes m}) -
{(E(\varrho^{\otimes m})^{\otimes n})/ n}| \leq \epsilon$. But, by
definition,
$(\varrho^{\otimes m})^{\otimes n} = \varrho^{\otimes mn}$ where the
equality relates equivalent density matrices on products of
isomorphic
local spaces.  Thus
$n \geq N$ implies
\[ \left|E(\varrho) - {E(\varrho^{\otimes m})\over m}\right|
\leq \left|E(\varrho) - {E(\varrho^{\otimes mn})\over mn}\right| +
\left|{E(\varrho^{\otimes mn})\over mn} -
{E(\varrho^{\otimes m}) \over m}\right| \leq
2\epsilon.
\]

(E4) follows.  The same proof shows the equivalence of (P4$'$) and
(P4). \hfill $\blacksquare$

\begin{lem}
\label{lemU}
Let $E$ be an entanglement measure which
satisfies (P1$'$), (P2), and (P4).  Then $E$ satisfies (P0) and
(P1).  Moreover, if $E$ is defined on mixed states and satisfies
either (E2) or (E6$'$), then (E0) is satisfied.
\end{lem}
{\bf Proof:} First we deal with separable states.

Choose $\epsilon > 0$.  Any pair of separable pure states are
interconvertible by local unitary operators.  If $\sigma$ is such a
state, then so is $\sigma^{\otimes n}$, and so, by (P2),
$E(\sigma) =  E(\sigma^{\otimes n})$.  But (P4) implies that
$E(\sigma) = E(\sigma^{\otimes n})/n$ and hence
$E(\sigma) = 0$.  This gives (P0) and the $d = 1$ case of
(P1).

Now let $\varrho \in \Sigma({\mathcal{H}}^A \otimes
{\mathcal{H}}^B)$ be a mixed separable state.  Expanding the states
$\varrho^{A}_i$ and $\varrho^{B}_i$ of Equation (\ref{e1}) into pure
components shows that $\sigma$ is a convex combination of pure
separable states: $\sigma = \sum_{i} p_{i} \sigma_i$.

Thus (E6$'$) is sufficient to go from (P0) to (E0).
But (E2) is also sufficient, because if $\Lambda_i:
{\mathcal{T}}({\mathcal{H}}^A \otimes {\mathcal{H}}^B) \rightarrow
{\mathcal{T}}({\mathcal{H}}^A \otimes {\mathcal{H}}^B)$ is a local
operation such that $\Lambda_i(\sigma_1) = \sigma_i$, then
$\Lambda := \sum_{i} p_{i} \Lambda_i$ is an \texttt{LQCC}
operation such that
$\Lambda(\sigma_1) = \sigma$ and so (E2) and (P0) yield
$E(\sigma) \leq E(\sigma_1) = 0$.

Now we turn to showing that, for
$d \geq 2$, $E(P^{d}_+) = \log_2 d$ follows from (P1$'$), (P2), and
(P4). By (P2$'$), $E(P^{d}_+)$ is independent of the
representative
of $P_+({\mathbb{C}}^d)$ considered.

Choose $\epsilon > 0$ and $d \geq 2$.  Choose $N > 1/\epsilon$.
Set $w(n) \equiv E(P^{n}_+)$.

By Nielsen's theorem, (P2) implies that
$w(d_1) \leq w(d_2)$ whenever $d_1 \leq d_2$.

Up to local isomorphisms, $(P^{d}_+)^{\otimes n} = P^{d^n}_+$, so
that,
by (P4),
$w(d) = w(d^n)/n$ for all $n$ and, by (P1$'$), $w(2) = w(2^n)/n=1$.

Choose $n_1, n_2 > N$ such that $2^{n_2 + 1} \geq d^{n_1} \geq
2^{n_2}$.  Then $\log_2 d \geq n_2/n_1$, $|{n_2 / n_1} - \log_2 d|
\leq
1/n_1 < \epsilon$ and, using (P4),
\[ |w(d) - \log_2 d| \leq
|w(d^{n_1}) - n_2|/n_1 + |{n_2 / n_1} - \log_2 d| \leq  |w(d^{n_1})
-
n_2 w(2)|/n_1 + \epsilon
\]
and
\[ |w(d^{n_1}) - n_2 w(2)|/n_1
=|w(d^{n_1}) - w(2^{n_2})|/n_1
\leq |w(2^{n_2 +1}) - w(2^{n_2})|/n_1 = 1/n_1  \leq
\epsilon. \]
It follows that $w(d)$ is arbitrarily close to $\log_2 d$. \hfill
$\blacksquare$

\begin{lem}
\label{lemE5}
(E5$'$) implies (E5$''$).  Indeed, (E5$'$) implies that
${E(\varrho^{\otimes m}) \over m} \to \inf
\left\{{E(\varrho^{\otimes m}) \over m} : m \geq 1 \right\}$.
\end{lem}
{\bf Proof:} (see \cite{Walters82} Theorem 4.9). Fix $k>0$. Every $m
\geq 1$
can be written $m = nk + r $ with $0 \leq r < k$. Then
for all $m>0$ set $f(m) := E(\varrho^{\otimes m})$.
(E5$'$) implies that
\[ {f(m) \over m} \leq { n f(k) + f(r) \over nk + r } \leq
{nf(k) \over nk} + {f(r) \over
nk} = {f(k) \over k} + {f(r) \over nk}.
\]
As $m \to \infty$ then $n \to \infty$ so
$\limsup_{m \to \infty} {f(m) \over m} \leq
{f(k) \over k}$ and thus $\limsup_{m \to \infty}
{f(m) \over m} \leq
\inf_{k
\geq 1}{f(k) \over k}$. Now $\inf_{k \geq 1}{f(k) \over k} \leq
\liminf_{m \to \infty} {f(m)
\over m}$ shows that $\lim_{m \to \infty}
{f(m) \over m}$ exists and equals $\inf_{m \geq 1}
{f(m) \over m}$. \hfill $\blacksquare$

\begin{prop} \label{reg}
Let $E$ be an entanglement measure which satisfies (E5$''$).  Then
\begin{itemize}
\item[(1)] $E^\infty$ satisfies (E4)
\item[(2)] If $E$ satisfies (E0), then so does $E^\infty$.
\item[(3)] If $E$ satisfies (E1), then so does $E^\infty$.
\item[(4)] If $E$ satisfies (E2), then so does $E^\infty$.
\item[(5)] If $E$ satisfies (E5), then so does $E^\infty$.
\item[(6)] If $E$ satisfies (E5) and (E6), then so does $E^\infty$.
\end{itemize} \end{prop}
\noindent\textbf{Proof}:

\noindent 1)  For all $m$ and $\varrho$,
\[ {E^{\infty}(\varrho^{\otimes m}) \over m} =
\lim_{n\rightarrow\infty} {E(\varrho^{\otimes nm})\over
nm} = E^{\infty}(\varrho). \]

\noindent 2) \ If $\sigma$ is separable, then so is $\sigma^{\otimes
n}$
for all $n$.

\noindent 3) \ If $P^d_+$ is a representative of
$P_+({\mathbb{C}}^d)$, then $(P^d_+)^{\otimes n}$ is a
representative
of $P_+({\mathbb{C}}^{d^n})$.

\noindent 4) \ If $\Lambda$ is \texttt{LQCC}, then so is
$\Lambda^{\otimes n}$
and
$\Lambda(\sigma)^{\otimes n} = \Lambda^{\otimes n}(\sigma^{\otimes
n})$.

\noindent 5) \ For all $\varrho, \sigma$ and $k \geq 1$, (E5)
implies that
\[ {E((\varrho \otimes \sigma)^{\otimes k})
\over k}
\leq {E(\varrho^{\otimes k}) \over k} +
{E(\sigma^{\otimes k}) \over k}. \]

\noindent 6) \  Suppose that $E$ satisfies (E5) and (E6).  Let
$\varrho, \sigma \in \Sigma({\mathcal{H}}^A \otimes
{\mathcal{H}}^B)$
and choose  $x_1, x_2 \in [0, 1]$ with $x_1 + x_2 = 1$.  Let
$\omega = x_1 \varrho + x_2 \sigma$.  Expanding $\omega^{\otimes n}$
as a sum of products, using convexity of $E$, and then using local
isomorphisms to re-order the terms in each product, gives
\[ E(\omega^{\otimes n}) \leq \sum_{k = 0}^n {n \choose k} x_1^k
x_2^{n-k} E( \varrho^{\otimes k} \otimes \sigma^{\otimes (n - k)})
\leq
\sum_{k = 0}^n {n \choose k} x_1^k x_2^{n-k}( E( \varrho^{\otimes
k})
+ E(\sigma^{\otimes (n - k)}))\]
where the second inequality is a consequence of (E5).  To complete
the
proof, we need the following lemma:

\begin{lem}
\label{lemE52} As $n \to \infty$,  ${1 \over n}\sum_{k = 0}^n {n
\choose k} x_1^k x_2^{n-k} E(\varrho^{\otimes k}) \to x_1
E^\infty(\varrho)$ and
${1 \over n}\sum_{k = 0}^n {n \choose k} x_1^k x_2^{n-k}
E(\sigma^{\otimes (n - k)}) \to x_2 E^\infty(\sigma)$.
\end{lem}
\noindent\textbf{Proof}:  It is sufficient to prove the first limit.
Set $g(m) = E(\varrho^{\otimes m})/m$ and $L =
E^\infty(\varrho)$.  Choose $\epsilon > 0$.  By Lemma \ref{lemE5}, there
exists
$K$ such that $k \geq K$ implies $|g(k) - L| < \epsilon/2$ and
there is a constant $C > 0$ such that $|g(k) - L| < C$ for all
$k$.
$N > K$ implies that
\[ {1 \over N}\sum_{k = 0}^K {N \choose k} k x_1^k x_2^{N-k} \leq
{K \over N}\sum_{k = 0}^N {N \choose k} x_1^k x_2^{N-k} = {K \over
N}.
\]

Set $h(x) = (x + y)^n = \sum_{k = 0}^n {n \choose k} x^k y^{n - k}$.
$x h'(x) = n x (x + y)^{n-1} = \sum_{k = 0}^n {n \choose k} k x^k
y^{n
- k}$.  Thus $x_1 + x_2 = 1$ implies that $\sum_{k = 0}^n {n
\choose k} k x_1^k x_2^{n-k} = n x_1$.

Choose $N_0 > K$ such that $K C/ N_0 < \epsilon/2$.  Then $N > N_0$
implies
\begin{eqnarray*} \left|{1 \over N}\sum_{k = 0}^N {N
\choose k} x_1^k x_2^{N-k} E(\varrho^{\otimes k}) -
x_1
E^\infty(\varrho) \right| &=& \left|{1 \over N}\sum_{k = 0}^N
{N \choose k} k x_1^k x_2^{N-k} g(k) - x_1
L \right| \\ &=& \left|{1 \over N}\sum_{k = 0}^N {N
\choose k} k x_1^k x_2^{N-k}(g(k) - L) \right| \\
&\leq & {1 \over N}\sum_{k = 0}^K {N \choose k} k x_1^k x_2^{N-k} C
\\ & & + {1 \over N}\sum_{k = K + 1}^N {N
\choose k} k x_1^k x_2^{N-k}(g(k) - L) \\
&\leq & K C/N + \epsilon/2 \sum_{k = K + 1}^N {N
\choose k} x_1^k x_2^{N-k} \\ & \leq & \epsilon.
\end{eqnarray*}
\hfill $\blacksquare$ \quad $\blacksquare$

Continuity (E3) is not mentioned in Proposition \ref{reg}, although
we could use Lemma \ref{lemE5} to deduce upper-semicontinuity from
(E3) and (E5$'$), as the infimum of a family of real continuous
functions is upper-semicon\-tinuous.  For an example which may be
relevant, consider the sequence of functions on $[0, 1]$ defined by
$f_n(x) = n x^n$.  Clearly $f_{m+n}(x) \leq f_m(x) + f_n(x)$.
$g_n(x)
= x^n$ converges (pointwise) as $n \to \infty$ to a discontinuous,
but
upper-semicontinuous, function.

%%%%%%%%%%%%%%%%%%%%%%%%%%%%%%%%%%%%%%%%%%%%%%%%%%%%%%%%%%%%%%%%%%%%%%%%%%

\section{Examples of important entanglement measures}
\label{sec-examples}

%%%%%%%%%%%%%%%%%%%%%%%%%%%%%%%%%%%%%%%%%%%%%%%%%%%%%%%%%%%%%%%%%%%%%%%%%%

In this section we will present some important entanglement measures
and check which of the postulates from Section \ref{sec-measures}
they
satisfy.

\subsection{Operational measures}
Here we shall describe two entanglement measures,
\emph{entanglement of distillation}  and \emph{entanglement cost}
\cite{BennettVSW97} (see also \cite{Rains_bound,Terhal}),
which are defined in terms of specific state conversions.

\begin{lem} \label{lemExist}  Let $\varrho \in
\Sigma({\mathcal{H}}^A\otimes {\mathcal{H}}^B)$ with
${\mathcal{H}}^A
\equiv {\mathcal{H}}^B \equiv {\mathcal{H}}$ and $\dim{\mathcal{H}}
=
d$.  Let $|\phi \rangle = |\phi^A \rangle \otimes |\phi^B \rangle
\in
{\mathcal{H}}^A\otimes {\mathcal{H}}^B$ be a separable wavefunction and
$P^d_+$ be a representative of $P_+({\mathbb{C}}^{d})$ on
${\mathcal{H}}^A \otimes {\mathcal{H}}^B$.
Then there exist \texttt{LQCC} operations $\Lambda_1$ and
$\Lambda_2$ such that $\Lambda_1(\varrho) = |\phi\rangle \langle \phi|$ and
$\Lambda_2(P^d_+) = \varrho$.
\end{lem}

\noindent\textbf{Proof}:  Let $(\psi^{A}_i)_{i = 1}^d$ (resp.
$(\psi^{B}_i)_{i = 1}^d$) be an orthonormal basis for
${\mathcal{H}}^A$
(resp. ${\mathcal{H}}^B$) and define $\Lambda_1$ by
\begin{eqnarray*}
\Lambda_1(\sigma) &\equiv& \sum_{j=1}^d \left(1^A \otimes
\left|\phi^B \right\rangle
\left\langle \psi^B_j \right|\right) \left(\sum_{i=1}^d
\left( \left|\phi^A \right\rangle
\left\langle \psi^A_i \right|
\otimes 1^B \right) \sigma \left(\left|\psi^A_i \right\rangle
\left\langle \phi^A \right|
\otimes
1^B \right) \right) \left(1^A \otimes \left|\psi^B_j
\right\rangle \left\langle
\phi^B\right| \right) \\
&=& \sum_{i,j=1}^d \left|\phi^A \otimes \phi^B \right\rangle
\left\langle \psi^A_i \otimes
\psi^B_j \right| \sigma \left| \psi^A_i \otimes \psi^B_j
\right\rangle \left\langle \phi^A
\otimes \phi^B \right| = |\phi\rangle \trace(\sigma) \langle \phi| =
|\phi\rangle \langle \phi|
\end{eqnarray*}
for all $\sigma \in \Sigma({\mathcal{H}}^A\otimes {\mathcal{H}}^B)$.

For $\Lambda_2$, we note that if $|\Psi\rangle\langle \Psi|
\in \Sigma({\mathcal{H}}^A\otimes {\mathcal{H}}^B)$ is any pure
state,
then, by Nielsen's theorem, there exists an \texttt{LQCC}
operation mapping
$P^d_+$ to $|\Psi\rangle\langle \Psi|$ because the distribution $({1
\over d})_{i=1}^d$ is majorized by any probability distribution on
$\{1, \dots, d\}$.  Now, as in the proof of Lemma \ref{lemU}, we can
construct $\Lambda_2$ as a convex combination of operations mapping
$P^d_+$ to pure components of $\varrho$. \hfill $\blacksquare$
\\

Given a state $\varrho$ on ${\mathcal{H}}^A\otimes {\mathcal{H}}^B$,
consider a sequence of $\texttt{LQCC}$ operations
$(\Lambda_n)$ with $\Lambda_n:
{\mathcal{T}}(({\mathcal{H}}^A)^{\otimes n} \otimes
({\mathcal{H}}^B)^{\otimes n}) \rightarrow
{\mathcal{T}}(({\mathcal{H}}^A)^{\otimes n} \otimes
({\mathcal{H}}^B)^{\otimes n})$.
Suppose, that
$\sigma_n\equiv\Lambda_n(\varrho^{\otimes n})$ satisfies
\begin{equation} \nonumber
\Vert P^{d_n}_+ - \sigma_n \Vert_1 \rightarrow 0
\end{equation}
for some representative $P^{d_n}_+$ of $P_+({\mathbb{C}}^{d_n})$ on
$({\mathcal{H}}^A)^{\otimes n} \otimes ({\mathcal{H}}^B)^{\otimes
n}$.
We call such a sequence
$(\Lambda_n)$ an
\texttt{LQCC}
\emph{distillation protocol}.
The asymptotic ratio
attainable via this protocol is then defined by
\begin{equation} \label{eq16}
E_D((\Lambda_n),\varrho) \equiv \limsup_{n\rightarrow \infty}
{\log_2 d_n \over n}.
\end{equation}

Lemma \ref{lemExist} shows that, for any state, a distillation
protocol always exists with $d_n \equiv 1$.

\begin{de} \label{distDef} The \emph{distillable
entanglement} or \emph{entanglement of distillation} $E_D$
is defined as the supremum of Equation (\ref{eq16})
over all possible \texttt{LQCC} distillation protocols:
\begin{equation}
E_D(\varrho) \equiv \sup_{(\Lambda_n)}E_D((\Lambda_n),\varrho).
\end{equation}
\end{de}

By construction $E_D$ satisfies the properties (E2) and
(E4) of entanglement measures. The proof is analogous to the proof
of
Lemma 1 in \cite{Terhal}. It is not known whether $E_D$ satisfies
(E3)
or (E6). (Indeed, as already mentioned,
there is evidence that (E6) may not be satisfied \cite{Shor2000}).
We
shall confirm in Lemma \ref{lemDCvN} that (E0) and (E1) are
satisfied.

The so-called \emph{entanglement cost} $E_C$ is
defined in a complementary way.
Given a state $\varrho$ consider a sequence of \texttt{LQCC}
operations $\Lambda_n : \mathcal{T}({\mathbb{C}}^{d_n}
\otimes {\mathbb{C}}^{d_n}) \to
\mathcal{T}(({\mathcal{H}}^A)^{\otimes n} \otimes
({\mathcal{H}}^B)^{\otimes n})$ transforming a representative of
$P_+({\mathbb{C}}^{d_n})$  into a state $\sigma_n$
such that
\begin{equation}
\Vert \sigma_n - \varrho^{\otimes n} \Vert_1 \to 0.
\nonumber \end{equation}
The asymptotic ratio attainable via this \emph{formation}-protocol
is
then given by
\begin{equation} \label{eq13}
E_C((\Lambda_n),\varrho) \equiv \liminf_{n \to \infty}
{\log_2 d_n \over n}.
\end{equation}

Once again Lemma \ref{lemExist} shows that, for any state, a
formation protocol always exists with $d_n \equiv d^n$ where $d =
\max\{\dim{\mathcal{H}}^A, \dim{\mathcal{H}}^B\}$.
\begin{de} \label{costDef}
The \emph{entanglement cost} $E_C$
is defined as the infimum of Equation (\ref{eq13})
over all possible \texttt{LQCC} formation protocols:
\begin{equation}
E_C(\varrho) \equiv \inf_{\{\Lambda_n\}} E_C((\Lambda_n),\varrho).
\end{equation}
\end{de}

By construction $E_C$ satisfies property (E2). As we shall discuss
in the next section, by \cite{Terhal} and Proposition \ref{reg},
it also satisfies (E0), (E1), (E2), (E4), (E5), and (E6). It is not
known whether it satisfies (E3). We shall also prove below that for
pure states both $E_D$ and $E_C$ are equal to the reduced von Neumann
entropy given by Equation (\ref{vN}). (This was first realized in
\cite{BBPS} and a rigorous proof was sketched in \cite{Nielsen99}.)

\subsection{Abstract measures}
The entanglement measures discussed in this subsection
quantify entanglement mathematically but their definitions do not
admit
a direct operational interpretation in terms of entanglement
manipulations.
The first one is the so-called {\it entanglement of formation}
\cite{BennettVSW97} which is
defined as follows:

\begin{de} Let ${\mathcal{H}}^A$ and ${\mathcal{H}}^B$ be finite
dimensional Hilbert spaces and let $\vert \psi \rangle \in
{\mathcal{H}}^A \otimes {\mathcal{H}}^B$, then the {\em entanglement
of
formation} is defined for pure states as
\addtocounter{equation}{1}
\alpheqn
\begin{equation}
E_F(P_\psi) := S_{\mathrm{vN}}(P_\psi),
\end{equation}
where $S_{\mathrm{vN}}(P_\psi)$ (defined in Equation (\ref{vN})) is
the von Neumann entropy of  either of the reduced density matrices
of $\vert \psi \rangle$. For mixed
states
$\varrho
\in {\Sigma}({\mathcal{H}}^A \otimes {\mathcal{H}}^B)$ we define
\begin{equation}
E_F(\varrho) := \inf \sum_i p_i E_F(P_{\psi_i})
\end{equation} \reseteqn
where  the infimum is taken over all possible decompositions of
$\varrho$ of the form
$\varrho=\sum_ip_i|\psi_i\ra\la\psi_i|$ with $p_i \geq 0$ for all
$i$
and $\sum_i p_i =1$.
\end{de}

The entanglement of formation satisfies (E0) -- (E3), (E5), and
(E6). In particular, (E2) was shown in Ref.~\cite{BennettVSW97},
(E3) in Ref.~\cite{Nielsen-cont}, and (E0),
(E1), (E5), and (E6) follow directly from the definition of
$E_F$.

The entanglement of formation $E_F$ is believed but not known
to be equal to the entanglement cost $E_C$.
However, it is known that the
{\it regularized} entanglement of formation $E^\infty_F$ (which
exists
by (E5$'$)), is equal to the entanglement cost \cite{Terhal}.  This
allows us to apply Proposition \ref{reg} to $E_C$.

Let us now present another important measure, namely, the
\emph{relative entropy
of entanglement} \cite{VedralPRK97,PlenioV98}. It is defined as
follows
\begin{equation}
E_{R}(\varrho) \equiv \inf_{\sigma}
S_{\mathrm{rel}}(\varrho|\sigma),
\label{Rains-measure}
\end{equation}
where $S_{\mathrm{rel}}(\varrho|\sigma) \equiv
\trace\varrho\log_2\varrho-\trace\varrho
\log_2 \sigma$ is
the quantum relative entropy, and where the
infimum is taken over all separable states $\sigma$.
One can consider variations
of the above measure, by changing the set of states over which the
infimum is taken
(this set should be
closed under {\texttt{LQCC}} operations though).
Like the entanglement of formation, $E_R$ satisfies (E0)--(E3),
(E5),
and (E6). In particular, (E1) and (E2) were shown in
Ref.~\cite{VedralPRK97}, (E3) in Ref.~\cite{DonaldH99},
(E0) follows immediately and (E5) almost immediately from the
definition of $E_R$, (E6) follows from the convexity of the quantum
relative entropy $S_{\mathrm{rel}}$.

The properties of $E_R$ and Proposition \ref{reg} show that the
\emph{regularized} relative entropy of entanglement $E^\infty_R$
exists and satisfies (E0), (E1), (E2), (E4), (E5), and (E6).  It
is shown in \cite{VollbrechtW00} that $E_R$ does \emph{not} satisfy
(E4).  This implies, of course, that $E_R$ and $E^\infty_R$ are not
always equal (cf.~\cite{Virmani}).

Finally, let us note that for pure states
both the entanglement of formation (by definition) and
the relative entropy of entanglement (as shown in \cite{PlenioV98},
\cite{Plenio00}) are equal to the reduced von Neumann entropy
$S_{\mathrm{vN}}$ (defined in Equation (\ref{vN}) above).  An
immediate consequence of the additivity of $S_{\mathrm{vN}}$ is that
$E^\infty_F = E_C$ and $E^\infty_R$ are also equal to
$S_{\mathrm{vN}}$ on pure states (see also Theorem \ref{unique}).

%%%%%%%%%%%%%%%%%%%%%%%%%%%%%%%%%%%%%%%%%%%%%%%%%%%%%%%%%%%%%%%%%%%%%%%%%

\section{Entanglement of distillation and entanglement cost  as
extreme measures}%

%%%%%%%%%%%%%%%%%%%%%%%%%%%%%%%%%%%%%%%%%%%%%%%%%%%%%%%%%%%%%%%%%%%%%%%%%

\label{sec-extreme}
In this section we improve the theorem
of Ref. \cite{Horodecki99} by giving precise
conditions under which $E_D$ and $E_C$ are lower and upper
bounds for entanglement measures.  We propose
three versions of the theorem.

\begin{prop} \label{propnn}
Suppose that $E$ is an entanglement measure defined on mixed states
which satisfies (E1)--(E4).
Then for all states $\varrho \in {\Sigma}({\mathcal{H}}^A
\otimes {\mathcal{H}}^B)$
\begin{equation}
E_D(\varrho) \leq E(\varrho) \leq  E_C(\varrho).
\end{equation} \end{prop}
\noindent \textbf{Proof}:  Choose $\epsilon > 0$.
We shall prove the result in three steps:

\noindent \textbf{I.}  First we prove that, having if necessary
passed
to a subsequence, there exists an integer $N_1 >0$ such that $n \geq
N_1$ implies
\begin{equation}
{E(\varrho^{\otimes n})\over n} \geq
E_D(\varrho)- \epsilon.
\label{es1}
\end{equation}
Consider a near-optimal \texttt{LQCC} protocol $(\Lambda_{n})_n$.
By the definition of distillable entanglement, there exists a
\texttt{LQCC} protocol $(\Lambda_{n})_n$ such that, after possibly
passing to a subsequence,
\addtocounter{equation}{1}
\alpheqn
\begin{equation}
\left\Vert P^{d_n}_+ - \Lambda_n(\varrho^{\otimes n})
\right\Vert_1 \to 0
\end{equation}
and
\begin{equation}
\left| E_D(\varrho) - {\log_2 d_{n} \over n} \right|
\leq \epsilon/2 \end{equation} \reseteqn
for all $n \geq N'_1$.
(E3) implies that
\begin{equation} \left|{E(\Lambda_n(\varrho^{\otimes n})) -
E\left(P^{d_n}_+ \right) \over 1 + n \log_2 d} \right| \to 0
\end{equation} as $n \to \infty$ where $d =
\dim {\mathcal{H}}^A \otimes {\mathcal{H}}^B$.  It follows that we
can
choose $N''_1 >0$ such that $n
\geq N''_1$ implies
\begin{equation}
\left|{E(\Lambda_n(\varrho^{\otimes n}))\over n} -
{E\left(P^{d_n}_+\right)\over n}\right| \leq \epsilon/2
\label{es3} \end{equation}
and so, using (E2), for $n \geq N_1 = \max\{N'_1, N''_1\}$,
\begin{equation}
{E(\varrho^{\otimes n})\over n} \geq {E(\Lambda_n(\varrho^{\otimes
n}))\over n} \geq {E \left(P^{d_n}_+ \right)
\over n}- \epsilon/2 ={\log_2 d_{n} \over n} -
\epsilon/2
\geq E_D(\varrho)- \epsilon.
\end{equation}

\noindent \textbf{II.}  As a second step, we prove that, having if
necessary passed to another (perhaps disjoint) subsequence, there
exists
an integer $N_2 \geq N_1$ such that $n
\geq N_2$ implies
\begin{equation}
{E(\varrho^{\otimes n})\over n} \leq
E_C(\varrho) + \epsilon.
\label{es2}
\end{equation}
This is similar to the first step.  Consider a
near-optimal protocol $(\Lambda_{n})_n$ for $\varrho$. We have
(after possibly passing to a suitable subsequence of
$(\Lambda_n)_n$),
for all sufficiently large $n$,
\begin{equation}
{E(\varrho^{\otimes n})\over n}   \leq
{E\left(\Lambda_{n}\left(P^{d_n}_+\right)\right) \over n} +
\epsilon/2
\leq
{E\left(P^{d_n}_+\right) \over n} + \epsilon/2
=  {\log_2 d_{n} \over
n} + \epsilon/2   \leq  E_C(\varrho) + \epsilon.
\end{equation}

\noindent \textbf{III.}  The final step is to invoke (E4) to give
\begin{equation} \label{es4} E_D(\varrho) - \epsilon
\leq E(\varrho) = {E(\varrho^{\otimes n})\over n} \leq E_C(\varrho)
+ \epsilon. \end{equation}
\hfill $\blacksquare$

Unfortunately, as we do not at present know of any function for
which
we can prove that postulates (E1)--(E4) hold for all states, it is
possible that
Proposition \ref{propnn}
may be empty.
Nevertheless, by modifying the final step of the proof, we can
obtain
the following:
\begin{prop} \label{propnn2}
Let $E$ be an entanglement measure defined on mixed states and
satisfying (E1), (E2), (E3), and (E5$''$).
Then for all states $\varrho \in
{\Sigma}({\mathcal{H}}^A \otimes {\mathcal{H}}^B)$,
\begin{equation}
E_D(\varrho) \leq E^{\infty}(\varrho) \leq  E_C(\varrho).
\end{equation}
\end{prop}
\textbf{Proof}: Without using condition (E4) or any properties of
$E^\infty$ except its existence, we can maintain the structure of
the previous proof, simply by replacing $E(\varrho)$ in (\ref{es4})
by
$E^\infty(\varrho)$.
\hfill
$\blacksquare$ \\

Proposition \ref{propnn2}
is certainly non-empty. Indeed, as mentioned in the
previous section, both the entanglement of formation and the
relative entropy of
entanglement satisfy all assumptions of the proposition. We obtain
\begin{co}
\label{EDleqEC}
The entanglement of distillation $E_D$
is less than or equal to the entanglement cost $E_C$
for all states.
\end{co}

Although, in physical terms, Corollary \ref{EDleqEC} seems almost
necessary, a rigorous proof requires some control both over changes
in
state and over changes in dimension.

Let us now consider  yet another version, where we weaken the
assumptions in the theorem on extreme measures of
Ref.~\cite{Horodecki99}. We impose the condition (E3$'$) which
is stronger than (P3) but weaker than (E3).

One mechanism for deriving condition (E3$'$) for a given function $E$ might
be to establish
the
inequalities
\begin{equation} \label{ineq}
f(\varrho)\leq E(\varrho)\leq g(\varrho)
\end{equation}
where $f,g$ are functions satisfying (E3$'$) which coincide
on pure states. We will take $f(\varrho) \equiv
S(\varrho_A)-S(\varrho)$ and
$g(\varrho) \equiv S(\varrho_A)$
(where   $\varrho_A:=\trace_{{\mathcal H}_B}\varrho$).
Both of these functions $f$ and $g$ do satisfy (E3$'$). This follows
immediately from two facts:
\begin{itemize}
\item[(i)] Fannes inequality \cite{Fannes,OhyaP93}
\begin{equation} |S(\sigma) - S(\varrho)| \leq \Vert \sigma -
\varrho \Vert_1 \log_2 \dim{\mathcal{H}} +
\eta(\Vert \sigma - \varrho \Vert_1)
\label{Fannes2}
\end{equation}
which holds for any two states $\sigma$ and $\varrho$ acting on
the Hilbert space $\mathcal{H}$ and satisfying
$\Vert \sigma-\varrho \Vert_1 \leq {1\over3}$;
here $\eta(s) \equiv - s \log s$ and $S$ denotes the standard von
Neumann entropy as above;
\item[(ii)]
$\Vert \sigma_A-\varrho_A\Vert_1 \leq \Vert \sigma
-\varrho\Vert_1$
where $\sigma_A$ and $\varrho_A$ are the
reduced density operators of $\varrho$ and $\sigma$ respectively.
\end{itemize}

With the above choices for $f$ and $g$
one can show that $E_F$ and $E_R$ satisfy the inequalities
in (\ref{ineq})
see \cite{BennettVSW97,PlenioV98,PMH98,Plenio00}.  Then,
$E_R^\infty$ and $E_F^\infty$ also satisfy inequalities
(\ref{ineq}),
because the additivity of the von Neumann entropy implies that both
$f$ and $g$ satisfy (E4). $E_D$ also satisfies
the inequality $E_D\leq g$ but we do not know whether or not it
satisfies the second inequality. However,
a stronger inequality (the so-called {\it hashing inequality}),
which would have many interesting implications, was conjectured
in Ref.~\cite{Horodecki00}. Strong evidence for this
conjecture was collected there.

We shall also use the weak form of convexity (E6$'$).
\begin{prop}
\label{propnn3}
Let $E$ be an entanglement measure defined on mixed states and
satisfying (E1), (E2), (E3$'$), and (E6$'$). Then
for all states $\varrho \in
{\Sigma}({\mathcal{H}}^A \otimes {\mathcal{H}}^B)$ we have
\begin{equation}
E_D(\varrho) \leq E(\varrho) \leq  E_C(\varrho)
\end{equation}
if (E4) holds and
\begin{equation}
E_D(\varrho) \leq E^\infty(\varrho) \leq  E_C(\varrho)
\end{equation}
if (E5) holds.
\end{prop}
{\bf Proof:} Step I of the proof of Proposition \ref{propnn} goes
through with (E3$'$) replacing (E3) in inequality (\ref{es3}).

To replace step II, we use the estimate $E_C\geq E_F^\infty$.  This
follows from Proposition \ref{propnn2} (but also of course from
Ref.~\cite{Terhal} where it was shown that $E_C=E_F^\infty$).
For any state $\varrho$ consider its finite decompositions
into pure states
\[
\varrho=\sum_ip_i|\psi_i\rangle\langle\psi_i|
\]
for which
\[
E_F(\varrho)=\sum_i p_i S_{{\mathrm{vN}}}(P_{\psi_i}).
\]
In Ref.~\cite{Uhlmann} it was shown that such a decomposition
exists.

As (E1)=(P1) $\Rightarrow$ (P1$'$), (E2) $\Rightarrow$ (P2), and
(E3$'$) $\Rightarrow$ (P3), we can apply Theorem
\ref{unique} below to show that
$E(P_{\psi_i})=S_{{\mathrm{vN}}}(P_{\psi_i})$ if $E$ satisfies (E4) and
$E^\infty(P_{\psi_i})=S_{{\mathrm{vN}}}(P_{\psi_i})$ if $E$ satisfies
(E5).

Now (E6$'$) implies, in the first case, that
$E(\varrho)\leq E_F(\varrho)$ (cf.~\cite{Uhlmann}) and hence
\[ E(\varrho) = {E(\varrho^{\otimes n})\over n}\leq {E_F(\varrho^{\otimes
n})\over n} \]
which yields the required upper bound when $n \to \infty$. For the second
case, we can use the proof of part (6) of Proposition \ref{reg} to show that
(E6$'$) holds for
$E^\infty$.  This yields
$E^\infty(\varrho)\leq E_F(\varrho)$ and
\[  E^\infty(\varrho) = {E^\infty(\varrho^{\otimes n})\over
n}\leq {E_F(\varrho^{\otimes n})\over n}. \]
Again the required bound follows on taking $n \to \infty$.
\hfill $\blacksquare$ \\

%%%%%%%%%%%%%%%%%%%%%%%%%%%%%%%%%%%%%%%%%%%%%%%%%%%%%%%%%%%%%%%%%%%%%%%%%%%

\section{The uniqueness theorem for entanglement measures}
\label{sec-uniqueness}

%%%%%%%%%%%%%%%%%%%%%%%%%%%%%%%%%%%%%%%%%%%%%%%%%%%%%%%%%%%%%%%%%%%%%%%%%%%

\begin{theo} \label{unique}
Let $E$ be a functional on pure states.  Then the following are
equivalent
\begin{itemize}
\item[(1)] $E$ satisfies (P1$'$), (P2),
(P3), and (P4$'$).
\item[(2)] $E$ satisfies (P0), (P1), (P2), (P3), and (P4).
%\item[(4)] $E$ satisfies (P0), (P1), (P2), (P3), and (P4$^0$).
\item[(3)] $E$ coincides with the reduced von Neumann
entropy $E= S_{{\mathrm{vN}}}$.
\end{itemize}
On the other hand, if $E$ satisfies (P0), (P1), (P2), and (P3), then $E$
satisfies (P5$''$) and, on pure states, $E^\infty = S_{{\mathrm{vN}}}$.
\end{theo}
\noindent\textbf{Proof}: The equivalence of (1) and (2) is proved in
Lemmas \ref{lemadd} and \ref{lemU}.
%It is clear that (4) implies (2).

It is clear that the reduced von Neumann
entropy satisfies (P0), (P1) and (P4). (P3) follows from the facts
(i)
and (ii) of the previous section. Finally (P2) is a consequence of
Nielsen's Theorem and the fact that the von Neumann entropy is a
Schur-concave function \cite{Nielsen00}.  Indeed, with the inductive
decomposition of \texttt{LQCC} operations introduced in our proof of
Nielsen's
theorem, we can prove (P2) just by showing, in the notation of
Equation (\ref{eqNiel}), that
$S_{{\mathrm{vN}}}(\Lambda(|\Psi\rangle \langle
\Psi|)) \leq S_{{\mathrm{vN}}}(|\Psi\rangle \langle \Psi|)$.
This amounts to
proving that, for $p_J \geq p_K$ and suitable $\delta$,
\[- (p_J + \delta) \log_2 (p_J + \delta) - (p_K - \delta) \log_2
(p_K - \delta) \leq - p_J \log_2 p_J - p_K \log_2 p_K\]
and this is easily confirmed by differentiating with respect to
$\delta$.

Now suppose that $E$ satisfies (P0), (P1), (P2), and (P3).  Using
(P2$'$), we may assume that ${\mathcal{H}}^A \equiv {\mathcal{H}}^B
\equiv {\mathcal{H}}$.  Suppose that $\dim{\mathcal{H}}=d$ and
let $\vert \psi \rangle \in {\mathcal{H}} \otimes {\mathcal{H}}$.
Write ${\mathtt{S}} \equiv S_{{\mathrm{vN}}}(\vert \psi \rangle
\langle \psi \vert)$ for the von Neumann entropy of the reduced
density
matrix of $\vert \psi \rangle$. Consider $n$ copies
of the wavefunction $\vert \psi \rangle$:
$\left\vert \psi^{\otimes n} \right\rangle
\in {\mathcal{H}}_{{\mathrm{tot}}} \equiv
{\mathcal{H}}^{\otimes n} \otimes {\mathcal{H}}^{\otimes n}$.   Let
$\{ q_j : j = 1, \dots, d\}$ be the set of eigenvalues of the
reduced density matrix of $\left\vert \psi \right\rangle$ and
$\{ p_i : i = 1, \dots, d^{2n} \}$ be the set of
eigenvalues of the reduced density matrix of $\left\vert
\psi^{\otimes
n}\right\rangle$.  Again using (P2$'$), we may adjust $d$ so that
$q_j
> 0$ for $j = 1, \dots, d$.  In view of (P0), we may also assume
that
${\mathtt{S}} > 0$.  Considered as a probability distribution,
$\{ p_i\}$ is the distribution for $n$ independent trials each with
distribution $\{ q_j \}$.  Choose bases  $(e_i) \subset
{\mathcal{H}}^{\otimes n}$ and $(f_i)\subset{\mathcal{H}}^{\otimes
n}$
such that
\[ \left\vert \psi^{\otimes n}
\right\rangle = \sum_i \sqrt{p_i} |e_i \rangle \otimes | f_i
\rangle.
\]
Choose $\epsilon>0$. By the asymptotic equipartition theorem
(\cite{CoverT91} Theorem 3.1.2), there exists an integer
$N \equiv N(\epsilon)$ such that, for all $n \geq N$, one can find a subset
$\mathtt{TYP} \equiv \mathtt{TYP}(n,\epsilon)$
of the set of indices $\{ i\}_{i=1}^{d^{2n}}$ with
the following properties:
\addtocounter{equation}{1}
\alpheqn
\begin{equation}
2^{-n({\mathtt{S}} + \epsilon)} \leq p_i \leq 2^{-n({\mathtt{S}} -
\epsilon)}, \quad \mbox{for } i \in {\mathtt{TYP}},
\label{eq1}
\end{equation}
\begin{equation}
p \equiv \sum_{i \in {\mathtt{TYP}}} p_i \geq 1-\epsilon,
\end{equation}
\begin{equation}
\# {\mathtt{TYP}} \leq 2^{n({\mathtt{S}} + \epsilon)}.
\label{eq3}
\end{equation} \reseteqn
Here $\# {\mathtt{TYP}}$ denotes the number of elements in
${\mathtt{TYP}}$.

Introduce another wavefunction $\vert \phi_n \rangle
\in {\mathcal{H}}_{{\mathrm{tot}}}$ given by
\[
\vert \phi_n \rangle \equiv {1\over \sqrt{p}} \sum_{i \in
{\mathtt{TYP}}}
\sqrt{p_i} | e_i \rangle \otimes |f_i \rangle.
\]
This wavefunction satisfies
\begin{equation}
\left| \left\langle \psi^{\otimes n} | \phi_n \right\rangle
\right|^2
= p
\geq
1-\epsilon \label{eq4}
\end{equation}
and so
\begin{equation}
\left\Vert \vert \psi^{\otimes n} \rangle
\langle \psi^{\otimes n} \vert - \vert \phi_n \rangle
\langle \phi_n \vert \right\Vert_1 = 2\sqrt{(1 -  \left|
\left\langle
\psi^{\otimes n} | \phi_n \right\rangle
\right|^2)} \leq  2\sqrt{\epsilon}. \label{uniq5}
\end{equation}

Now, the crucial observation (cf.~\cite{Nielsen99}) is that
for $\epsilon < \min\{ {1 \over
2}{\mathtt{S}}, {1 \over 2}\}$ and $n$ sufficiently large, there
exist completely positive maps $\Lambda_n$ and $\Lambda_n'$ such
that
\addtocounter{equation}{1}
\alpheqn
\begin{equation} \label{eq21a}
\Lambda_n(\left \vert \phi_n \right\rangle \left\langle \phi_n
\right\vert)= P^{a}_+
\end{equation}
for $P^a_+$ a representative of $P_+({\mathbb{C}}^a)$ in
${\mathcal{H}}_{{\mathrm{tot}}}$ with
$\left|{\log_2 a\over n} - {\mathtt{S}} \right| < \epsilon + {2
\over
n}$ and
\begin{equation} \label{eq21b}
\Lambda'_n(P^{b}_+) =
\left \vert \phi_n \right\rangle \left\langle \phi_n \right\vert
\end{equation}
\reseteqn
for $P^{b}_+$ a representative of $P_+({\mathbb{C}}^{b})$ in
${\mathcal{H}}_{{\mathrm{tot}}}$ with $\left|{\log_2 b\over n} -
{\mathtt{S}} \right| < \epsilon + {1 \over n}$. Indeed, to see
Equation
(\ref{eq21a}), set $a \equiv \lfloor p 2^{n({\mathtt{S}}-\epsilon)}
\rfloor$, i.e., $a$ is the largest integer smaller than or equal to
$p 2^{n({\mathtt{S}}-\epsilon)}$.  Then $a \leq p
2^{n({\mathtt{S}}-\epsilon)} \leq p/p_i$ and we see
that the distribution $\left( {p_i\over p} \right)_{i \in
{\mathtt{TYP}}}$  is majorized by $\left( {1\over a}
\right)_{i=1}^a$,
hence Equation (\ref{eq21a}) follows from Nielsen's Theorem.
Equation (\ref{eq21b}) follows by a similar argument when we
take $b\equiv \lceil p 2^{n({\mathtt{S}}+\epsilon)} \rceil$, i.e.,
$b$ is the smallest integer larger than or equal to $p
2^{n({\mathtt{S}}+\epsilon)}$.  The conditions on $\epsilon$ and $n$ are
sufficient to go from $a \equiv \lfloor p 2^{n({\mathtt{S}}-\epsilon)}
\rfloor$ to $\left|{\log_2 a\over n} - {\mathtt{S}} \right| < \epsilon + {2
\over n}$ and from $b\equiv \lceil p 2^{n({\mathtt{S}}+\epsilon)} \rceil$ to
$\left|{\log_2 b\over n} - {\mathtt{S}} \right| < \epsilon + {1 \over n}$,
ensuring, for example, that $a \ne 0$.

Now choose a sequence $( \epsilon_j )_{j \in \mathbb{N}}$ of
positive numbers such that $\epsilon_j\rightarrow0$ for $j \to
\infty$. Suppose that $( n_k )_{k \in \mathbb{N}}$ is a sequence of integers
such that $n_k \to \infty$ and $\frac{E(\vert \psi^{\otimes n_k} \rangle
\langle \psi^{\otimes n_k} \vert)}{n_k} \to L$ for some $L$.

For each $j$, choose $n_{k_j} \geq \max\{N(\epsilon_j),
1/\epsilon_j\}$.  We can apply the postulates (P0)--(P3)
to obtain the following estimates:
\begin{eqnarray*}
\frac{E(\vert \psi^{\otimes n_{k_j}} \rangle
\langle \psi^{\otimes n_{k_j}} \vert)}{n_{k_j}}
& = & \frac{E(\vert \psi^{\otimes n_{k_j}} \rangle
\langle \psi^{\otimes n_{k_j}} \vert) -
E(\vert \phi_{n_{k_j}} \rangle \langle \phi_{n_{k_j}} \vert)}{n_{k_j}}
+ \frac{E(\vert \phi_{n_{k_j}} \rangle \langle \phi_{n_{k_j}}
\vert)}{n_{k_j}}
\\
& \geq & \frac{E(\vert \psi^{\otimes n_{k_j}} \rangle
\langle \psi^{\otimes n_{k_j}} \vert) -
E(\vert \phi_{n_{k_j}} \rangle \langle \phi_{n_{k_j}} \vert)}{n_{k_j}}
+ \frac{E(\Lambda_{n_{k_j}}(\vert \phi_{n_{k_j}}
\rangle \langle \phi_{n_{k_j}} \vert))}{n_{k_j}}
\\ & = & \frac{E(\vert \psi^{\otimes n_{k_j}} \rangle
\langle \psi^{\otimes n_{k_j}} \vert) -
E(\vert \phi_{n_{k_j}} \rangle \langle \phi_{n_{k_j}} \vert)}{n_{k_j}}
+ \frac{E \left(P^{a_{n_{k_j}}}_+ \right)}{n_{k_j}} \\
& = & \frac{E(\vert \psi^{\otimes n_{k_j}} \rangle
\langle \psi^{\otimes n_{k_j}} \vert) -
E(\vert \phi_{n_{k_j}} \rangle \langle \phi_{n_{k_j}} \vert)}{n_{k_j}}
+ \frac{\log_2 a_{n_{k_j}}}{n_{k_j}}.
\end{eqnarray*}
As $j \to \infty$, the first term vanishes due to (P3) and the
second approaches $S_{\mathrm{vN}}(P_\psi)$ (cf.~Ref.~\cite{Nielsen-cont}).
This implies that
$L  \geq
S_{\mathrm{vN}}(\vert
\psi \rangle \langle \psi \vert)$.
The proof of the inequality $L
\leq S_{\mathrm{vN}}(\vert
\psi \rangle \langle \psi \vert)$ is
similar:
\begin{eqnarray*}
\frac{E(\vert \psi^{\otimes n_{k_j}} \rangle
\langle \psi^{\otimes n_{k_j}} \vert)}{n_{k_j}}
& = & \frac{E(\vert \psi^{\otimes n_{k_j}} \rangle
\langle \psi^{\otimes n_{k_j}} \vert) -
E(\vert \phi_{n_{k_j}} \rangle \langle \phi_{n_{k_j}} \vert)}{n_{k_j}}
+ \frac{E(\vert \phi_{n_{k_j}} \rangle \langle \phi_{n_{k_j}}
\vert)}{n_{k_j}}
\\
& \leq &  \frac{E(\vert \psi^{\otimes n_{k_j}} \rangle
\langle \psi^{\otimes n_{k_j}} \vert) -
E(\vert \phi_{n_{k_j}} \rangle \langle \phi_{n_{k_j}} \vert)}{n_{k_j}}
+ \frac{E \left(P^{b_{n_{k_j}}}_+\right)}{n_{k_j}} \\
& = & \frac{E(\vert \psi^{\otimes n_{k_j}} \rangle
\langle \psi^{\otimes n_{k_j}} \vert) -
E(\vert \phi_{n_{k_j}} \rangle \langle \phi_{n_{k_j}} \vert)}{n_{k_j}}
+ \frac{\log_2 b_{n_{k_j}}}{n_{k_j}}.
\end{eqnarray*}

We have now shown that every limit point of the sequence $\frac{E(\vert
\psi^{\otimes n} \rangle \langle \psi^{\otimes n} \vert)}{n}$ has the value
$L=S_{\mathrm{vN}}(|\psi\rangle\langle\psi|)$.  But, by (P1), (P2), and Lemma
\ref{lemExist}, this sequence is bounded, and so (P5$''$) holds with
$E^\infty(|\psi\rangle\langle\psi|) = L =
S_{\mathrm{vN}}(|\psi\rangle\langle\psi|)$.  This proves the final statement
of the theorem.  On the other hand, if (P4) holds then
$L=E(|\psi\rangle\langle\psi|)$, and so we have proved that (2) implies
(3). This completes the proof of Theorem \ref{unique}. \hfill
$\blacksquare$
\\

It is natural to wonder whether the conditions in Theorem
\ref{unique} can be weakened, and, in particular, whether (P3) is
necessary. That it is has been noted by Vidal \cite{Vidal00}.
Consider the entanglement measures defined on pure states by
$S_\infty(\sigma) = -\log_2 p_1(\sigma)$ where
$p_1(\sigma)$ is the largest coefficient in a Schmidt decomposition
of $\sigma$ and by $S_0(\sigma) = \log d(\sigma)$ where $d$ is the
number of non-zero coefficients.  $S_0$ and $S_\infty$ both satisfy (P0),
(P1), (P2) (by Nielsen's theorem), and (P4).  $S_\infty$ is even trace
norm
continuous on Hilbert spaces of fixed dimension.  (P3) however does
not hold for either.  This is, of course, a consequence of Theorem
\ref{unique}.  An explicit example of the failure of (P3) for $S_\infty$
is
provided by the states
$\sigma_n \equiv \vert \Psi_n \rangle \langle \Psi_n \vert$,
$\varrho_n \equiv \vert \Phi_n \rangle \langle \Phi_n \vert$ with
Schmidt
decompositions $\vert \Psi_n \rangle \equiv \sqrt{1 \over 2^n}
\vert \psi_1
\psi_1 \rangle + \sum_{i = 2}^{4^n - 2^n + 1} {1 \over 2^n} \vert \psi_i
\psi_i \rangle$  and $\vert \Phi_n \rangle \equiv \sum_{i = 1}^{4^n}
{1
\over 2^n} \vert \psi_i \psi_i \rangle$ for some orthonormal family
$(\vert \psi_i \rangle)$ of wavefunctions.
In fact, any entanglement measure $E$
defined on pure states and satisfying (P0), (P1), (P2), and (P4),
will
satisfy $S_\infty(\sigma) \leq E(\sigma) \leq S_0(\sigma)$ for all pure
$\sigma$.  The upper bound here is a consequence of Lemma
\ref{lemExist} while, for the lower bound, we modify
the proof of Theorem \ref{unique} using the fact that $\vert
\psi^{\otimes n} \rangle \langle \psi^{\otimes n} \vert$ can always be
converted without approximation into $P^c_+$ where $c$ is the largest
integer smaller than or equal to $1/p_1$.

An example of a measure on pure states satisfying (P0), (P1), (P2), (P3),
but not (P4), is given by
$E(\sigma) = 2(1 - p_1(\sigma)) S_{\mathrm{vN}}(\sigma)$ for $p_1(\sigma)
\geq {1
\over 2}$, $E(\sigma) = S_{\mathrm{vN}}(\sigma)$ for $p_1(\sigma) \leq {1
\over 2}$.

Finally, let us consider entanglement of distillation and
entanglement cost in the above context. Using the maps constructed
in Theorem \ref{unique}, we show that they are equal to
$S_{\mathrm{vN}}$.  We have already noted that for $E_C$ this also
follows from \cite{Terhal}.

\begin{lem} \label{lemDCvN}
The entanglement of distillation $E_D$ and the entanglement cost
$E_C$ both coincide on pure states with the von Neumann
reduced entropy $E_D(P_\psi) = E_C(P_\psi)  =
S_{\mathrm{vN}}(P_\psi)$ for all $\vert \psi  \rangle \in
{\mathcal{H}} \otimes {\mathcal{H}}$.
\end{lem}
\textbf{Proof}:
{}From Section \ref{sec-extreme} we know that $E_D \leq E_C$.
It suffices
to show that on pure states $E_D\geq S_{\mathrm{vN}}$
and $E_C\leq S_{\mathrm{vN}}$.
We will continue to use the notation from the proof of Theorem
\ref{unique}.

That $E_C(P_\psi)\leq S_{\mathrm{vN}}(P_\psi)$ follows directly from
the definition of $E_C$, using the operations defined by the
$\Lambda'_{n_j}$ which satisfy Equation (\ref{eq21b}) and estimate
(\ref{uniq5}).

To show that $E_D(P_\psi)\geq S_{\mathrm{vN}}(P_\psi)$, let us apply
the map $\Lambda_{n_j}$ from Equation (\ref{eq21a}) to the state
$\vert \psi^{\otimes {n_j}} \rangle \langle \psi^{\otimes {n_j}}
\vert$.
We only need check that the resulting state
$\Lambda_{n_j} \left(
|\psi^{\otimes {n_j}}\ra\la\psi^{\otimes {n_j}}| \right)$
approaches $P_+^{a}$ as $j\to\infty$.  But, by Lemma \ref{cont},
\[  \left\Vert \Lambda_{n_j} (
|\psi^{\otimes {n_j}}\ra\la\psi^{\otimes {n_j}}|) - P_+^{a}
\right\Vert_1
= \left\Vert \Lambda_{n_j} (
|\psi^{\otimes {n_j}}\ra\la\psi^{\otimes {n_j}}|) - \Lambda_{n_j}
(|\phi_{n_j} \rangle \langle \phi_{n_j} |) \right\Vert_1 \leq
\left\Vert
|\psi^{\otimes {n_j}}\ra\la\psi^{\otimes {n_j}}| - |\phi_{n_j}
\rangle
\langle \phi_{n_j} | \right\Vert_1 \]
and once again estimate (\ref{uniq5}) is sufficient. \hfill
$\blacksquare$ \\

\noindent With the results obtained in this paper, we can now
prove that $E_D$ is convex on pure decompositions, i.e.,

\begin{lem}
\begin{equation}
E_D \left( \sum_i p_i | \psi_i \rangle \langle \psi_i | \right)
\leq \sum_i p_i E_D \left(|\psi_i\rangle \langle \psi_i| \right),
\end{equation} where $p_i \geq 0$ for all $i$ and $\sum_i p_i =1$.
\end{lem}
\noindent\textbf{Proof}:  We have seen that $E_C$ is
convex and satisfies  $E_D \leq E_C$.  Using Lemma \ref{lemDCvN}
gives
\begin{eqnarray}
E_D \left( \sum_i p_i \vert \psi_i \rangle \langle \psi_i \vert
\right) & \leq &
E_C \left(\sum_i p_i \vert \psi_i \rangle \langle \psi_i \vert
\right)
\nonumber \\ & \leq &
\sum_i p_i E_C \left(|\psi_i\ra\la\psi_i| \right)
= \sum_i p_i S_{{\mathrm{vN}}}(P_{\psi_i})
= \sum_i p_i E_D \left( \vert \psi_i \rangle \langle \psi_i \vert
\right).
\end{eqnarray}
\hfill $\blacksquare$


\begin{thebibliography}{99}
\bibitem{BennettVSW97} C.H.~Bennett, D.P.~DiVincenzo, J.A.~Smolin,
and W.K.~Wootters,
Mixed-state entanglement and quantum error correction,
Phys.~Rev.~A \textbf{54}, 3824 %-3851
(1996), quant-ph/9604024.
\bibitem{PlenioV98} V.~Vedral and M.B.~Plenio,
Entanglement measures and purification procedures,
{Phys.~Rev.~A}\textbf{57}, 1619 %-1633
(1998),
quant-ph/9707035.
\bibitem{Vidal00} G.~Vidal,
Entanglement monotones,
{Journ.~Mod.~Opt.~}\textbf{47}, 355 %-376
(2000),
quant-ph/9807077.
\bibitem{Horodecki99} {M.~Horodecki, P.~Horodecki and R.~Horodecki,
}
Limits for entanglement measures,
{Phys.~Rev.~Lett.~}\textbf{84},
2014 %-2017
(2000),
quant-ph/9908065
\bibitem{PopescuR97} S.~Popescu, D.~Rohrlich,
Thermodynamics and the measure of entanglement,
Phys.~Rev.~A
\textbf{56}, R3319 %-R3321
(1997),
quant-ph/9610044.
\bibitem{Nielsen-cont}
M.A.~Nielsen,
Continuity bounds for entanglement,
Phys.~Rev.~A \textbf{61}, 64301 (2000),
quant-ph/9908086.
\bibitem{extreme-finite} An analogous result for the finite regime
was obtained in Ref.~\cite{Vidal00}.
\bibitem{Shor2000}
P.W.~Shor, J.A.~Smolin and B.M.~Terhal,
Nonadditivity of bipartite distillable entanglement follows
from conjecture on bound entangled Werner states,
Phys.~Rev.~Lett.~\textbf{86}, 2681 (2001),
quant-ph/0010054.
\bibitem{DonaldH99} M.J.~Donald and M.~Horodecki,
Continuity of relative entropy of entanglement,
Phys.~Lett.~A \textbf{264},
257 %-260
(1999),
quant-ph/9910002.
\bibitem{VollbrechtW00} K.G.H.~Vollbrecht and R.F.~Werner,
Entanglement measures under symmetry,
Phys.~Rev.~A \textbf{64}, 062307 (2001),
quant-ph/0010095.
\bibitem{Linden}
N.~Linden, S.~Popescu, B.~Schumacher and M.~W.~Westmoreland,
Reversibility of local transformations of multiparticle entanglement,
quant-ph/9912039.
\bibitem{Rudolph01} O.~Rudolph, A uniqueness theorem for
entanglement measures, J.~Math.~Phys.~\textbf{42}, 2507 (2001),
quant-ph/0105104.
\bibitem{Rudolph00}
O.~Rudolph,
A new class of entanglement measures,
J.~Math.~Phys.~\textbf{42}, 5306 (2001),
math-ph/0005011.
\bibitem{Schmidt}
E.~Schmidt, Zur Theorie der linearen und nichtlinearen
Integralgleichungen. I.~Teil: Entwicklung willk\"urlicher
Funktionen nach Systemen vorgeschriebener,
Math.~Annalen \textbf{63}, 433 (1907).
\bibitem{Kraus71} K.~Kraus,
General state changes in quantum theory,
{Ann.~Phys.~(N.Y.)} \textbf{64}, 311 %-335
(1971).
\bibitem{Davies76} E.B.~Davies, \emph{Quantum Theory of Open
Systems}, Academic, London, 1976.
\bibitem{Kraus83} K.~Kraus, \emph{States, Effects and
Operations}, Springer-Verlag, Berlin, 1983.
\bibitem{Choi75} M.-D.~Choi,
Completely positive linear maps on complex matrices,
Lin.~Alg.~Appl.~\textbf{10}, 285 %-290
(1975).
\bibitem{dhrv1} M.J.~Donald, M.~Horodecki, and O.~Rudolph,
The uniqueness theorem for entanglement measures, quant-ph/0105017
v1.
\bibitem{Rains_bound}
E.M.~Rains,
An improved bound on distillable entanglement,
Phys.~Rev.~A {\bf 60}, 179 (1999),
quant-ph/9809082.
\bibitem{nonlocality}
C.H.~Bennett, D.P.~DiVincenzo, C.A.~Fuchs, T.~Mor, E.~Rains,
P.W.~Shor, J.A.~Smolin and W.K.~Wootters,
Quantum nonlocality without entanglement,
Phys.~Rev.~A {\bf 59}, 1070 %-1091
(1999),
quant-ph/9804053.
\bibitem{Nielsen99} M.A.~Nielsen,
Conditions for a class of entanglement transformations,
Phys.~Rev.~Lett.~\textbf{83}, 436 %-439
(1999),
quant-ph/9811053.
\bibitem{Har99}
L.~Hardy,
Method of areas for manipulating the entanglement
properties of one copy of a two-particle pure entangled
state,
Phys.~Rev.~A {\bf 60}, 1912 %-1923
(1999), quant-ph/9903001.
\bibitem{JenSch00} J.G.~Jensen, R.~Schack, Simple algorithm for
local
conversion of pure states,
Phys.~Rev.~A \textbf{63}, 062303 (2001),
quant-ph/0006049.
\bibitem{BBPS}
C.~H.~Bennett, H.~J.~Bernstein, S.~Popescu and B.~Schumacher,
Concentrating partial entanglement by local operations, Phys.~Rev.~A {\bf
53}, 2046 (1996), quant-ph/9511030.
\bibitem{VedralPRK97} V.~Vedral, M.B.~Plenio, M.A.~Rippin and
P.L.~Knight,
Quantifying entanglement,
{Phys.~Rev.~Lett.~}\textbf{78}, 2275 %-2279
(1997),
quant-ph/9702027.
\bibitem{Joathan}
D. Jonathan and M. Plenio,
Minimal conditions for local pure-state entanglement manipulation,
Phys.~Rev.~Lett.~\textbf{83}, 1455 (1999),
quant-ph/9903054.
\bibitem{Walters82} P.~Walters, \emph{An Introduction to Ergodic
Theory}, Springer-Verlag, New York, 1982.
\bibitem{Terhal}
P.~Hayden, M.~Horodecki and B.~Terhal,
The asymptotic entanglement cost of preparing a quantum state,
J.~Phys.~A \textbf{34}, 6891 (2001),
quant-ph/0008134.
\bibitem{Uhlmann} A.~Uhlmann,
Entropy and optimal decompositions of states relative to a
maximal commutative subalgebra,
Open Sys.~\& Inf.~Dyn.~\textbf{5},
209 (1998),
quant-ph/9704017.
\bibitem{Fannes} M.~Fannes,
A continuity property of the entropy density for spin lattice
systems,
Commun.~Math.~Phys.~{\bf 31},
291 (1973).
\bibitem{OhyaP93} M.~Ohya and D.~Petz, \emph{Quantum Entropy and Its
Use}, Springer-Verlag, New York, 1993, p.~22.
\bibitem{PMH98}
P.~Horodecki, M.~Horodecki and R.~Horodecki,
Entanglement and thermodynamical analogies,
Acta Phys.~Slovaca {\bf 48}, 141
(1998),
quant-ph/9805072.
\bibitem{Virmani}
K.~Audenaert, J.~Eisert, E.~Jan\'e, M.B.~Plenio, S.~Virmani, and B.~De Moor,
Asymptotic relative entropy of entanglement,
Phys.~Rev.~Lett.~\textbf{87}, 217902 (2001),
quant-ph/0103096.
\bibitem{Plenio00}
M.B.~Plenio, S.~Virmani and P.~Papadopoulos,
Operator monotones, the reduction criterion and the relative
entropy,
J.~Phys.~A {\bf 33},
L193 (2000),
quant-ph/0002075.
\bibitem{Horodecki00}
M.~Horodecki, P.~Horodecki and R.~Horodecki,
Unified approach to quantum capacities: towards
quantum noisy coding theorem,
Phys.~Rev.~Lett.~{\bf 85}, 433 (2000),
quant-ph/0003040.
\bibitem{Nielsen00} M.A.~Nielsen,
Characterizing mixing and measurement in quantum mechanics,
Phys.~Rev.~A \textbf{63}, 022114 (2001),
quant-ph/0008073
\bibitem{CoverT91} T.M.~Cover and J.A.~Thomas, \emph{Elements of
Information Theory}, John Wiley, New York, 1991.
\end{thebibliography}
\end{document}